\documentclass[twoside]{article}
\usepackage{amsmath,amsfonts,amssymb}
\usepackage{qic,epsfig}
\usepackage{cite}
\usepackage{cases}

\renewcommand{\theequation}{\arabic{section}.\arabic{equation}}

\newtheorem{thm}{Theorem}
\newtheorem{cor}{Corollary}
\newtheorem{lem}{Lemma}
\newtheorem{defn}{Definition}

\renewenvironment{proof}{\par\noindent{\bf Proof.}}{$\quad\Box$\par}
\newcommand{\ket}[1]{| #1 \rangle}
\newcommand{\bra}[1]{\langle #1 |}

\begin{document}
\setlength{\textheight}{8.0truein}    

 \runninghead{Quotient Algebra Partition and Cartan Decomposition for $su(N)$ I}
            {Zheng-Yao Su}

\normalsize\textlineskip \thispagestyle{empty}
\setcounter{page}{1}

\vspace*{0.88truein}

\alphfootnote

\fpage{1}

\centerline{\bf Quotient Algebra Partition and Cartan
Decomposition for $su(N)$ I} \vspace*{0.035truein}
\centerline{\footnotesize Zheng-Yao Su\footnote{Email:
zsu@nchc.narl.org.tw
}} \centerline{\footnotesize\it National Center for
High-Performance
 Computing,}
 \centerline{\footnotesize\it National Applied Research Laboratories,
 Taiwan, R.O.C.}

\vspace*{0.21truein}

\abstracts{
 An algebraic structure, {\em Quotient Algebra Partition} or QAP, is introduced in
 a serial of articles. The structure QAP is universal to Lie Algebras and enables
 algorithmic and exhaustive Cartan decompositions.
 The first episode draws the simplest form of the structure in terms of
 the spinor representation.
 }{}{}

\vspace*{3pt} \vspace*{1pt}\textlineskip

\section{Introduction}\label{secintro}
\renewcommand{\theequation}{\arabic{section}.\arabic{equation}}
\setcounter{equation}{0}
 This serial of articles give a thorough exposition for the framework
 called {\em Quotient Algebra Partition (QAP)}.
 Originally discerned in 2005~\cite{OriginQAPSu} in the motivation of constructing Cartan decompositions~\cite{Khaneja1,Zhang,SBullock},
 a QAP is a partition over a Lie algebra
 that consists of abelian subspaces closed under commutation relations.
 Lie algebras and Lie groups are commonly known by
 their algebraic features from structures and geometric
 properties as applied to symmetric spaces~\cite{Helgason,Knapp}.
 While, the QAP structure particularly manifests their {\em combinatorial traits}~\cite{SuTsai1,SuTsai2,SuTsai3}.

 As an opening discussion slightly amended from the original version~\cite{OriginQAPSu},
 the 1st episode provides a simplest picture for the QAP structure in terms of the spinor representation.
 The structure of this form permits algorithmic and exhaustive Cartan decompositions of type {\bf AI}.
 In the successive episode~\cite{SuTsai1},
 the framework QAP is rephrased in the language of the {\em s-representation} for spinor generators,
 {\em cf.} Appendix~\ref{appsrep}.
 Thanks to this language,
 a QAP of the simplest form is established under a minimum number of conditions
 governed by a {\em bi-subalgebra} of rank zero, {\em i.e.}, a Cartan subalgebra.
 Besides, all Cartan subalgebras of $su(N)$ are classified
 and generated {\em shell by shell} through the process of the {\em subalgebra extension}.
 Thirdly~\cite{SuTsai2},
 a quotient algebra partition undergoes further partitions
 generated by bi-subalgebras of higher ranks.
 The {\em refined} versions of quotient algebra partitions of higher ranks allow not only Cartan decompositions of type {\bf AI}
 but also decompositions of types {\bf AII} and {\bf AIII},
 resorting to an operation called the {\em tri-addition}.
 A special type of transformations, referred to as {\em $s$-rotations},
 are introduced to attain the computational universality.
 It shows mappings preserving the QAP structure are composed of spinor-to-spinor $s$-rotations.
 In the final episode~\cite{SuTsai3},
 two kinds of partitions on unitary Lie algebras are created by  {\em nonabelian} bi-subalgebras,
 which reveals a {\em partition duality}.
 By applying appropriate coset rules,
 the two partitions return to a quotient algebra partition when the generating bi-subalgebra is {\em abelian}.
 Of significance is the universality of the QAP structure to classical and exceptional Lie algebras.
 A newly unveiled application of this framework~\cite{SuTsaiQAPFT}
 asserts that every action admits fault tolerance in every quantum code.
 This may pave a way for scalable fault tolerance quantum computation.

 Although most of examinations focus on $su(N)$ of a {\em nodal dimension},
 {\em i.e.,} $N=2^p$, the structure QAP is universally existing in
 an arbitrary Lie algebra of finite dimension.
 In this episode, the structure QAP is portrayed in the simplest form mainly
 in the {\em spinor} representation.
 Within a QAP, the algebra is partitioned into a certain number of abelian subspaces,
 and these subspaces form an algebra under an abelian and binary operation.
 Structured {\em binarily and combinatorially}, the framework QAP algorithmically produces exhaustive decompositions
 over a Lie algebra {\em level by level},
 and thus also resulted factorizations for the associated group actions.
 The presentation offers two views of a QAP structure,
 {\em i.e.},
 the quotient algebra and co-quotient algebra.

\section{Conjugate Partition}\label{secconjupar}
\renewcommand{\theequation}{\arabic{section}.\arabic{equation}}
\setcounter{equation}{0} \noindent
 To have a quick access to the structure of a quotient algebra, the exposition will begin
 with examples for the Lie algebras $su(6)$ and $su(8)$.
 Although the structure is representation
 independent, simply for convenience, 
 the generators of these algebras are written in tensor products
 of the Pauli and Gell-Mann matrices.

 The first step of introducing the quotient-algebra structure to the Lie algebra
 $su(N)$ is to select a {\bf \em center subalgebra} and let it be denoted by ${\cal A}$.
 In this preliminary episode, only the basic forms,
 quotient algebras of {\em rank zero}, will be concerned.
 To construct a such quotient algebra, an arbitrary Cartan subalgebra,
 a maximal abelian subalgebra, is taken from the algebra $su(N)$.
 With this choice of the center subalgebra, more abelian modules will be discerned.
 An abelian subalgebra of $7$ elements as listed in the central column of Fig.~\ref{stepsu8}
 is an example of the center subalgebra for $su(8)$,
 ${\cal A}=\{\sigma_{3}\otimes I\otimes I, I\otimes\sigma_{3}\otimes I,
            I\otimes I\otimes\sigma_{3}, \sigma_{3}\otimes\sigma_{3}\otimes I,
            \sigma_{3}\otimes I\otimes \sigma_{3}, I\otimes\sigma_{3}\otimes\sigma_{3},
             \sigma_{3}\otimes\sigma_{3}\otimes\sigma_{3}\}$.
 It may be easiest to form a Cartan subalgebra
 within a Lie algebra by collecting all its diagonal generators.
 The second step is 
 to assign an arbitrary generator $s\in su(N)-{\cal A}$ as a {\em seed},
 for example $ I\otimes I\otimes \sigma_{1}$ at the left hand side
 of the center subalgebra in Fig.~\ref{stepsu8}(a).
 Four other generators are produced at the right hand side
 by calculating the commutator of the seed with all elements
 of the center subalgebra, {\em i.e.}, 
 mapping the seed by the adjoint representation $ad_{\cal A}(s)=[s,{\cal A}]$.
 Then in the reverse step,
 three more generators are added to the LHS column as shown in Fig.~\ref{stepsu8}(b) after
 calculating the commutator $[g_0,{\cal A}]$, where $g_0$ is an arbitrary generator
 in the RHS column.

 A nice property is observed that both the {\em modules} of these two columns are abelian.
 Let $W_1$ and $\hat{W}_1$ denote the vector subspaces respectively spanned
 by the generators in the LHS and RHS columns and
 each of them is considered
 the {{\bf \em conjugate}} of the other, noting that $[W_1,\hat{W}_1]\subset \cal{A}$.
 Via the same procedure, more conjugate pairs of abelian subspaces
 $\{W_i,\hat{W}_i\}$
 are produced by taking a seed outside the center subalgebra
 and the existing conjugate pairs, from $\{W_{1},\hat{W}_{1}\}$
 to $\{W_{i-1},\hat{W}_{i-1}\}$, and calculating the required  commutators
 of the adjoint representation $ad_{\cal A}$.
 The exposition will temporarily be based on the current version of
 the {\em algorithm primitive}.

 Halting while no any seed generator being left, this procedure leads to a partition
 of the original algebra, for instance the one given in Fig.~\ref{csu8}.
 As stated in the following Lemma,
 there always exists a such {\bf \em conjugate partition} for
 the Lie algebra $su(N)$.
 \vspace{6pt}
 \begin{lem}\label{lemma}
 With an abelian subspace ${\cal A}$ suitably taken as
 the {\bf\em center subalgebra}, the Lie algebra $su(N)$ admits
 the {\bf\em conjugate partition} consisting of ${\cal A}$
 and a finite number $q$
 of {\bf\em conjugate pairs}
 $\{W_i,\hat{W}_i\}, 1\leq i\leq q$, namely,
 \begin{equation}\label{L1}
 su(N)= {\cal A} \oplus W_{1} \oplus \hat{W}_{1} \oplus\cdots\oplus W_{i} \oplus \hat{W}_{i}  \oplus
   \cdots \oplus W_{q} \oplus \hat{W}_{q},
 \end{equation}
 where
 the subspaces $W_i$ and $\hat{W}_i$ are abelian and obey the
 commutation relations, $\forall\hspace{2pt} 1\leq i\leq q$,
 \begin{equation}\label{L2}
 [W_{i},{\cal A}]\subset \hat{W}_{i}, \ \ [\hat{W}_{i},{\cal A}]\subset W_{i} \ \
 {\rm and} \ \ [W_{i},\hat{W}_{i}]\subset {\cal A}.
 \end{equation}
 \end{lem}
 \vspace{6pt}
 This lemma also provides the definition of the conjugate partition that is
 generally associated to quotient algebras of any ranks~\cite{SuTsai1,SuTsai2,SuTsai3}.
 The proof of this partition for quotient algebras of rank zero will be given
 in Theorem~\ref{thm01}.

 \section{Quotient Algebra}\label{secQA}
 \renewcommand{\theequation}{\arabic{section}.\arabic{equation}}
\setcounter{equation}{0} \noindent
 In addition to the conjugate partition composed of a certain number of abelian subspaces,
 of great interest is an algebraic structure embedded among these subspaces.
 Let a conjugate partition of $su(8)$ illustrated in Fig.~\ref{csu8} be taken as an example.
 It is easy to verify the fact that the commutator of any two elements from two subspaces of the partition
 must be either $\{0\}$ or included in a third subspace.
 This {\em closure} feature reveals the algebraic structure of thematic attention, 
 the {\bf {\em quotient algebra}}.
 \vspace{6pt}
 \begin{defn}\label{defn}
 For a conjugate partition generated by the center subalgebra ${\cal A}$
 and comprising a number $q$ of conjugate pairs
 of abelian subspaces $\{W_i,\hat{W}_i\}, 1\leq i\leq q$, the subspaces
 form a {\bf \em quotient algebra},
 denoted as a multiplet of partition
 $\{{\cal Q(A};q)\}\equiv\{{\cal A},W_i,\hat{W}_i;1\leq i\leq q\}$, if
  the {\bf\em condition of closure} is satisfied under the operation of the commutator:
 that is, 
 there always exists a unique third pair $\{W_k,\hat{W}_k\}$
 for two arbitrary conjugate pairs $\{W_i,\hat{W}_i\}$ and $\{W_j,\hat{W}_j\}$,
 $1\leq k\neq i,j \leq q$, such that
 \begin{equation}\label{D1}
 [W_i,W_j]\subset \hat{W}_k, \
 [W_i,\hat{W}_j]\subset W_k \ \ {\rm and} \ \
 [\hat{W}_i,\hat{W}_j]\subset \hat{W}_k;
 \end{equation}
 here the center subalgebra ${\cal A}$ acts as the identity of the operation
 within a conjugate pair,
 i.e.,
 $\forall\hspace{2pt} 1\leq i\leq q,$
\begin{equation}\label{D2}
 [W_{i},{\cal A}]\subset \hat{W}_{i}, \ \ [\hat{W}_{i},{\cal A}]\subset W_{i} \ \
 {\rm and} \ \ [W_{i},\hat{W}_{i}]\subset {\cal A}.
\end{equation}
 \end{defn}
 \vspace{6pt}
 In brief, a quotient algebra is a partition of a given algebra where
 each partitioned subspace is abelian and considered an individual module, 
 and more importantly these modules 
 are {\em closed} under
 the operation of the commutator following 
 the condition of closure.
 The number of conjugate pairs is essential, for which indicates the {\em rank}
 of a quotient algebra. Yet, the notation of the algebra
 allows the abbreviation $\{{\cal Q(A)}\}$ when the number specification is
 no needed.
 There exists the quotient-algebra structure not only in $su(2^p)$ but also in
 the Lie algebra $su(N)$ of dimension $N\neq 2^p$.
 As shown in Fig.~\ref{csu6} similarly, the center subalgebra of $5$ diagonal operators
 $\{I\otimes \sigma_3, \mu_3\otimes I, \mu_8\otimes I,\mu_3\otimes\sigma_3,
    \mu_8\otimes\sigma_3\}$ generates a quotient algebra in $su(6)$.
 Being a generating set of $su(3)$, the generators $\mu_j, j=1,2,\dots,8$,
 denote the Gell-Mann matrices, {\em ref}. Appendix~A.
 In this quotient algebra, the subspaces
 $W_i$ or $\hat{W}_i$ are spanned by either $2$ or $3$ generators, in contrast to equally
 $4$ in those of $su(8)$. To construct a conjugate partition and the corresponding
 quotient algebra, the center subalgebra is not restricted to a subspace of
 diagonal generators.
 Figures~\ref{ncsu8} and~\ref{ncsu6} present instances of
 quotient algebras for $su(8)$ and $su(6)$ acquired by taking
 other choices of center subalgebras. These instances foretell
 the following theorems.
 \paragraph{Main Theorem}
 {\em Every Lie algebra $su(N)$ 
 admits structures of quotient and co-quotient algebras up
 to rank $r_0$
 and its quotient and co-quotient algebras of rank $r$ 
 are respectively isomorphic to those of $su(2^p)$, 
 where $2^{p-1}<N\leq 2^p$, $0\leq r\leq r_0\leq p$,
 and the dimension has the factorization $N=2^{r_0}N'$
 with $N'$ being odd.} 
 \vspace{6pt}

 \noindent
 Similar to a quotient algebra, a co-quotient algebra~\cite{SuTsai2} is an algebraic structure
 realizing
 the conjugate partition and the condition of closure as prescribed in Definition~\ref{defn},
 which instead is generated
 by a center subalgebra quite different from that for a quotient algebra.
 The systematic exposition of quotient and co-quotient algebras
 requires the language of the {\em s-representation}, {\em ref}. Appendix~\ref{appsrep}.
 Phrased in the $s$-representation,
 these structures can be classified into different ``{\em ranks}" according
 to the kinds of their center subalgebras.
 The complete treatment of quotient and co-quotient algebras 
 shall be given in the $2$nd and the $3$rd episodes~\cite{SuTsai1,SuTsai2}, thus to which
 the proof of Main Theorem is postponed.
 At this beginning stage, only the simplest scenario,
 the quotient algebra of rank zero generated by a Cartan subalgebra, will be asserted.
 \vspace{6pt}
 \begin{thm}\label{thm01}
 Every Cartan subalgebra ${\cal A}$ of the Lie algebra $su(N)$, $2^{p-1}<N\leq 2^p$,
 can generate a quotient algebra of rank zero $\{{\cal Q(A};2^p-1)\}$.
 \end{thm}
 \vspace{3pt}
 \begin{proof}
 To prove this theorem along with Lemma~\ref{lemma}, the Lie algebra $su(2^p)$ of
 dimension $N=2^p$ will be considered first, and then followed by the extension
 to general dimensions as stated in Corollary~\ref{coro}.
 Rather than constructing the general form of a rank-zero quotient algebra with resort to
 the $s$-representation as 
 in the $2$nd episode~\cite{SuTsai1}, a straightforward
 proof in a simpler language will be presented here.

 For every Cartan subalgebra ${\cal A}\in su(N)$,
 there exists a transformation $U\in SU(N)$ to map ${\cal A}$ to
 its {\bf \em intrinsic coordinate} or {\bf \em eigenspace}, 
 where all elements of the subalgebra ${\cal A}$ are simultaneously diagonalized.
 Let the Cartan subalgebra consisting of only diagonal operators
 be regarded as the {\bf \em intrinsic Cartan subalgebra} or
 the {\bf \em intrinsic center subalgebra} and written in the reserved notation ${\cal C}$.
 In other words, owing to the connectedness of the group, there exists $U\in SU(N)$ for every ${\cal A}\in su(N)$,
 such that $U{\cal A}U^\dag={\cal C}$.
 Thus, it suffices to give the exposition mainly in the intrinsic coordinate.

 \paragraph{$\lambda$-Representation.}
 The vector space of the intrinsic center subalgebra ${\cal C}$ of
 $su(4)$ for example is
 spanned by the $3$ diagonal operators,
 $I\otimes\sigma_3=$ $diag\{1,-1,$ $1,-1\}$,
    $\sigma_3\otimes\sigma_3 =$ $ diag\{1,-1,$ $-1,1\}$  and
    $\sigma_3\otimes I =$ $ diag\{1,1,-1,-1\}$,
 or equivalently by another set of $3$ independent operators
    $diag\{1,-1,0, 0\}$,
    $diag\{1, 0,-1,$ $0\}$  and  $diag\{1,0,0,-1\}$.
 In general, it is of convenience to have a generating set of
 the intrinsic center subalgebra ${\cal C}$ of the Lie algebra $su(N)$ formed
 by a number $N-1$ of $N\times N$ diagonal operators: 
 $diag\{1,-1,$ $0,\cdots,0\}$, $diag\{1,0,-1,0,\cdots,0\}$,
 $\dots$ and $diag\{1,0,\cdots,0,-1\}$.
 According to the algorithm primitive, calculating the commutators of a
 seed generator with ${\cal C}$,
 or mapping the seed by the adjoint representation $ad_{\cal C}$, produces
 the generators in the conjugate subspace.
 To examine this operation, 
  the so-called
 {\bf $\lambda$-{\em representation}}, an extension of the well-known Gell-Mann matrices,
 is employed. 
 In this representation, a $\lambda$-generator $\lambda_{ij}$ is
 an off-diagonal $N\times N$ matrix and
 plays the role of $\sigma_1$, where
 the only two nonzero entries the $(i,j)$-th and $(j,i)$-th are both written to be $1$.
 Another $\lambda$-generator $\hat{\lambda}_{ij}$, being the {\em conjugate}
 of $\lambda_{ij}$ and an off-diagonal $N\times N$ matrix too,
 takes the role of $\sigma_2$ and has nonzero values only at the
 $(i,j)$-th and the $(j,i)$-th entries that are assigned with $-i$ and $i$ respectively.
 More details of this representation are referred to Appendix A.
 According to Eqs.~\ref{A1}-\ref{A3},
 the adjoint representation $ad_{\cal C}$ leaves a $\lambda_{ij}$
 or a $\hat{\lambda}_{ij}$ {\em invariant within the pair}.
 The $\lambda$-generators may well act as building blocks to 
 form the conjugate pairs of abelian subspaces.
 In terms of these generators, some {\em combinatorial traits} 
 will be read in the following construction of
 conjugate partitions and quotient algebras.

 \paragraph{Binary Partitioning.}
 A set of the $\lambda$-generators in an appropriate grouping can serve as a
 basis for an abelian subspace of a quotient algebra. Such a grouping
 implies a specific partition on the subscripts of the $\lambda$-generators.
 Since two of these generators commute as long as there having no repetition in their subscripts,
 a maximal abelian subspace in $su(N)-{\cal C}$ is spanned by
 a set of $N/2$ elements: \{$\lambda_{ij}$ or $\hat{\lambda}_{kl}$: the subscripts
 of the generators are assigned according to a partition of the integers
 from $1$ to $N$, $i,j,k,l=1,2,\dots,N$\}.
 For instance, two sets of $4$ generators
 $\{\lambda_{15},\hat{\lambda}_{26},\lambda_{37},\hat{\lambda}_{48}\}$ and
 $\{\lambda_{16},\hat{\lambda}_{28},\hat{\lambda}_{37},\hat{\lambda}_{45}\}$
 in $su(8)$
 respectively form an abelian subalgebra.
 Although both the partitions of generator subscripts lead to creating an abelian subspace,
 only the former kind fits the further use to hold
 a quotient algebra.
 Generators of conventional types bring in the clue.
 Take one spinor generator $\sigma_3\otimes I\otimes\sigma_1 \in su(8)$ for example,
 which reads as $\lambda_{12}+\lambda_{34}-\lambda_{56}-\lambda_{78}$
 in the $\lambda$-representation and is a vector in the subspace $W_1$ spanned
 by the $4$ generators $\lambda_{12},\lambda_{34},\lambda_{56},$ and $\lambda_{78}$.
 By exercising the algorithm primitive described above, the $4$ independent generators of
 the conjugate subspace $\hat{W}_1$
 are produced from the commutator $[\sigma_3\otimes I\otimes\sigma_1,{\cal C}]$.
 Conversely, the commutator $[\hat{g},{\cal C}]$,
 $\hat{g}$ being any one of the $4$ conjugate generators in $\hat{W}_1$,
 fills in the rest $3$ independent generators of
 the subspace $W_1$.
 These two elementary steps are in practice implementing Eqs.~\ref{A1} and~\ref{A2}.
 A critical observation is noted that the subscripts of
 the $\lambda$-generators so produced in the same conjugate pair share
 a common binary pattern of partitioning,
 here particularly termed as the {\bf \em binary partitioning}.
 For the subspace $W_1$, the subscripts of the $\lambda$-generators
 have the common pattern of {\em bit-wise addition} $i'+j'=001$,
 here $i'=i-1$ and $j'=j-1$ being written in their binary expressions.
 Since the mapping $ad_{\cal C}$ leaves
 a $\lambda$-generator invariant in its conjugate pair,
 all generators belonging to the same pair should retain
 the identical subscript pattern of the binary partitioning.
 That is, {\em the binary partitioning is an invariant under
 the adjoint representation} $ad_{\cal C}$.
 It is legitimate to {\em encode} such a subscript pattern into a binary string and
 attach the string to the conjugate pair.
 The subspaces $W_1$ and $\hat{W}_1$ are hence redenoted
 as $W_{001}$ and $\hat{W}_{001}$.

\paragraph{Conjugate Partition.}
 The continued step is to arbitrarily pick a spinor generator as a seed
 outside this conjugate pair and the center subalgebra.
 The example $\sigma_3\otimes\sigma_1\otimes\sigma_1=
          \lambda_{14}+\lambda_{23}-\lambda_{58}-\lambda_{67}$
 exhibits another pattern of the binary partitioning
 $i'+j'=011$ with the subscripts of the $\lambda$-generators $i'=i-1$ and $j'=j-1$ in
 their binary expressions. Then the conjugate pair
 $W_{011}$ and $\hat{W}_{011}$ are obtained
 by calculating the commutators guided in the algorithm primitive.
 Likewise, via this procedure the abelian subspaces,
 associated to the strings from $001$ to $111$, appear pair by pair, {\em ref}. Fig.~\ref{Gsu8}.
 These $7$ conjugate pairs along with the center subalgebra ${\cal C}$
 form a conjugate partition of $su(8)$.
 This procedure is generally applicable to the Lie algebra $su(2^p)$, of which
 every spinor generator in the $\lambda$-representation
 carries a certain subscript pattern of the binary partitioning.
 Let a binary string of $p$ digits encode a subscript pattern.
 Say associated to the string $\zeta\in Z^p_2$,
 an arbitrary spinor generator $g_\zeta\in su(2^p)$ is taken as a seed
 outside ${\cal C}$ and the existing conjugate pairs.
 With the aid of Eqs.~\ref{A1} and~\ref{A2},
 a new conjugate pair $\{W_{\zeta},\hat{W}_{\zeta}\}$ is created
 through the mappings $[g_\zeta,{\cal C}]$ and $[\hat{g}_\zeta,{\cal C}]$,
 where $\hat{g}_\zeta$ is a generator yielded from the former commutator.
 Such a step is recursively applied until no generator is left and
 a total number $2^p-1$ of conjugate pairs attached with
 distinct binary-partitioning strings all have appeared.
 Therefore, a conjugate partition, Eq.~\ref{L1}, completes in the Lie algebra $su(2^p)$.
 {\em The invariance of the binary partitioning within a conjugate pair}
 as formulated in Eq.~\ref{L2}   
 is a straightforward implication of Eqs.~\ref{A1}-\ref{A3}.

\paragraph{Quotient Algebra.}
 Most importantly, the binary partitioning
 encodes the structure of a quotient algebra.
 A such structure becomes manifest as being articulated
 in the phrasing of the binary partitioning.
 For three abelian subspaces
 associated to three binary-partitioning strings of $p$ digits
 $\zeta,\eta$ and $\xi$,
 the condition Eq.~\ref{D1} reads as
 $[W_{\zeta},W_{\eta}]\subset\hat{W}_{\xi}$,
 $[W_{\zeta},\hat{W}_{\eta}]\subset W_{\xi}$, and
 $[\hat{W}_{\zeta},\hat{W}_{\eta}]\subset\hat{W}_{\xi}$.
 According to Eqs.~\ref{A4}-\ref{A6}, the commutator of two
 $\lambda$-generators in $su(2^p)$
 with the subscripts $(i,j)$ and $(j,k)$, or $(k,j)$, contributes
 a $\lambda$-generator of the subscript $(i,k)$; here
 $i,j$ and $k$ are three distinct integers.
 Suppose that the subscript $(i,j)$ is associated to the string
 $\zeta$ and $(j,k)$ to $\eta$, and thus there have the
 identities of bit-wise additions $i'+j'=\zeta$ and $j'+k'=\eta$ for
 $i'=i-1$, $j'=j-1$ and $k'=k-1$ in their binary expressions.
 The string associated to the resulted subscript $(i,k)$
 is then coerced to be $\zeta +\eta$, for $j'+j'=0$.
 In this phrasing, the intrinsic Cartan subalgebra is in nature 
 an abelian subspace attached with the identity string, the string of all zeros,
 {\em i.e.}, ${\cal C}=W_{\bf 0}=\hat{W}_{\bf 0}$.
 It reaches the conclusion that the condition of closure Eq.~\ref{D1} as well as
 the conjugate partition Eq.~\ref{D2} can be pronounced
 in one elegant formulation, $\zeta,\eta\in Z^p_2$,
 \begin{equation}\label{D15}
 [W_{\zeta},W_{\eta}]\subset\hat{W}_{\zeta +\eta}, \
 [W_{\zeta},\hat{W}_{\eta}]\subset W_{\zeta +\eta}, \ \ {\rm and}\ \
 [\hat{W}_{\zeta},\hat{W}_{\eta}]\subset\hat{W}_{\zeta +\eta}.
 \end{equation}
 This formulation hints a feature of binary combinatorics in quotient algebras.
 An example for $su(8)$ is illustrated in Fig.~\ref{binarysu8}.

 The {\em intrinsic} quotient algebra of rank zero ${\cal \{Q(C};2^p-1)\}$ given by,
 actually ``partitioned by,"
 ${\cal C}\subset su(2^p)$ is therefore validated. The quotient algebra of rank zero
 ${\cal \{Q(A)\}}$ given
 by another Cartan subalgebra ${\cal A}$ is always transformable to the former one
 by a unitary action $U\in SU(2^p)$.
 Namely, the two quotient algebras are equivalent under a conjugation mapping
 ${\cal \{Q(A)\}}=U^{\dagger}{\cal \{Q(C)\}}U$
 provided $U{\cal A}U^{\dagger}={\cal C}$.

 \paragraph{Subscript Multiplication.}
 Apparently there are many more options of conjugate partitions.
 To fulfil the condition of Lemma~\ref{lemma} in $su(2^p)$, the binary partitioning is
 not a unique choice of the subscript arrangement.
 As aforementioned, a conjugate partition is achieved as long as
 the $\lambda$-generators 
 in an abelian subspace of every conjugate pair have no repetition in their subscripts.
 However, to further realize the structure of a quotient algebra, 
 the consistence in the subscripts, {\em i.e.}, following the binary partitioning
 or its equivalence, is necessary.
 Take an abelian subspace $V\subset su(8)$ for example that is spanned
 by the $4$ generators,
 $V=span\{\lambda_{16},\lambda_{25},\lambda_{37}.\lambda_{48}\}\subset W_{100}\cup W_{101}$.
 This grouping of subscripts is considered inconsistent due to bearing
 two different strings of the binary partitioning.
 As a consequence, a half of the $8$ generators produced from the commutator $[V,W_{011}]$
 fall in the subspace $W_{110}$ and another half in $W_{111}$.
 Specifically, the inconsistence causes the partition to violate the condition of closure.
 It will be shown immediately that, to accommodate in $su(2^p)$
 the structure of the intrinsic quotient algebra of rank zero given by ${\cal C}\subset su(2^p)$,
 the subscripts of the $\lambda$-generators are obliged to follow the
 binary partitioning or its permutations.

 For this purpose, the following demonstration is to reexamine
 how the condition of closure is satisfied
 under the operation of a commutator for two abelian subspaces
 in the intrinsic quotient algebra.
 Thanks to Eqs.~\ref{A4}-\ref{A6}, a commutator operation can
 be reduced to as simple as a {\em subscript multiplication} of
 two integer pairs:
 $(i,j)*(j,k)=(i,k)$, where $i,j$ and $k$ are the three different subscripts
 of the two $\lambda$-generators in a commutator and an integer pair here has no order,
 {\em i.e.,} $(i,j)=(j,i)$. The other multiplications are corresponding to the case
 either $\{0\}$ for two commuting generators, or the inclusion in 
 ${\cal C}$ for two generators
 belonging to one conjugate pair.
 The {\em subscript table} of $su(4)$
 is rendered as follows,
 \vspace{6pt}
 \begin{flushleft}
 $(1,2)\hspace{0.2cm}(3,4)$\\
 $(1,3)\hspace{0.2cm}(2,4)$\\
 $(1,4)\hspace{0.2cm}(2,3)$
 \end{flushleft}
 \vspace{6pt}

 \hspace{-0.55cm}which represents the $3$ conjugate pairs of a conjugate partition.
 Simply displaying the conjugate pairs of a conjugate partition,
 a subscript table is orderless in the sense
 that there is no order for integer pairs in one row and
 no order either for these rows in the table.
 Since the multiplication of two integer pairs from any two rows
 always falls in the rest row, this table represents a quotient algebra in $su(4)$ too.
 An easy check leads to the fact that all permutations of the $4$ integers 
 make the table unchanged,
 noting the orderless of a subscript table. It indicates that the binary partitioning
 is invariant with respect to the subscript permutation as the dimension $N=4$.

 Now proceed to the subscript table of $su(8)$,

 \vspace{6pt}
 \begin{flushleft}
 $(1,2) \hspace{0.2cm} (3,4) \hspace{0.2cm} (5,6) \hspace{0.2cm} (7,8) $ \\
 $(1,3) \hspace{0.2cm} (2,4) \hspace{0.2cm} (5,7) \hspace{0.2cm} (6,8) $ \\
 $(1,4) \hspace{0.2cm} (2,3) \hspace{0.2cm} (5,8) \hspace{0.2cm} (6,7) $ \\
 $(1,5) \hspace{0.2cm} (2,6) \hspace{0.2cm} (3,7) \hspace{0.2cm} (4,8) $.
 \end{flushleft}
 \vspace{6pt}

\hspace{-0.55cm}There are drawn only $4$ rows of integer pairs in the table
 to represent $4$ conjugate pairs.
 For according to the condition of closure,  the other $3$ rows of integer pairs
 are already decided by these $4$ rows. 
 This may be called the {\bf \em pre-decision rule} that, for a quotient algebra of $su(2^p)$,
 the other $2^p-p-1$ conjugate pairs are determined by a number $p$
 of {\em independent} pairs.
 Before the {\em interaction} row $\{(1,5)\ (2,6)\ (3,7)\ (4,8)\}$
 being taken into account,
 any permutations respectively exercised
 in the integer sets $\{1,2,3,4\}$ and $\{5,6,7,8\}$ are permitted,
 for these two sets are disjoint and have no ``{\em interaction}."
 Yet, some inconsistency may be introduced to the pairs
 while a such {\em partial} permutation is
 carried out along with the consideration of the interaction row .
 An example is as the table,
 \vspace{6pt}
 \begin{flushleft}
 $(1,2) \hspace{0.2cm} (3,4) \hspace{0.2cm} (5,7) \hspace{0.2cm} (6,8) $ \\
 $(1,3) \hspace{0.2cm} (2,4) \hspace{0.2cm} (5,6) \hspace{0.2cm} (7,8) $ \\
 $(1,4) \hspace{0.2cm} (2,3) \hspace{0.2cm} (5,8) \hspace{0.2cm} (6,7) $ \\
 $(1,5) \hspace{0.2cm} (2,6) \hspace{0.2cm} (3,7) \hspace{0.2cm} (4,8) $.
 \end{flushleft}
 \vspace{6pt}
 where the integers $6$ and $7$ are interchanged in the first $3$ rows,
 the {\em beginning} rows,  but no corresponding permutation
 is followed in the $4$th row, the {\em interaction} row.
 The order of a row here is referring
 to the display order in the table and has no meaning to the
 conjugate partition.
 This table is literally equivalent to
 \vspace{6pt}
 \begin{flushleft}
 $(1,2) \hspace{0.2cm} (3,4) \hspace{0.2cm} (5,6) \hspace{0.2cm} (7,8) $ \\
 $(1,3) \hspace{0.2cm} (2,4) \hspace{0.2cm} (5,7) \hspace{0.2cm} (6,8) $ \\
 $(1,4) \hspace{0.2cm} (2,3) \hspace{0.2cm} (5,8) \hspace{0.2cm} (6,7) $ \\
 $(1,5) \hspace{0.2cm} (2,7) \hspace{0.2cm} (3,6) \hspace{0.2cm} (4,8) $.
 \end{flushleft}
 \vspace{6pt}
 where the original integer pairs, complying with the binary partitioning,
 are preserved in the first $3$ rows, and the interchange of
 integers $6$ and $7$ instead takes place in the $4$th row.
 The condition of closure is violated when
 the interaction row is multiplied by either one
 of two beginning rows, the $1$st and the $2$nd, in the table.
 This instance is no atypical. Any permutation of integers in the
 {\em beginning} rows can be transferred to as a permutation
 in the {\em interaction} row once the binary partitioning is recovered
 in the former rows.
 Thus, whether the condition of closure is met is depending on
 if the integer pairs are consistently
 arranged in the interaction row.

 A recursive argument is developed for the general assertion.
 The subscript table of the Lie algebra $su(2^p)$ can be prepared by,
 in addition to one interaction row,
 {\em gluing} the two tables of $su(2^{p-1})$ in the way that
 each of the $2^{p-1}-1$ beginning rows is acquired by
 collecting integer pairs of the identical binary partitioning from
 the two tables. The integer pairs in one table are written in
 those from the $1$st set of $2^{p-1}$ integers
 $s_1=\{1,2,\dots,2^{p-1}\}$, 
 and the pairs in the other table are in integers from the $2$nd set
 $s_2=\{2^{p-1}+1,2^{p-1}+2,\dots,2^{p}\}$.
 The interaction row consists of $2^{p-1}$ integer pairs
 $(x,y)$ where $x\in s_1$ and $y\in s_2$.
 Suppose that
 there exist in the interaction row at least two integer pairs $(a,b)$ and $(c,d)$,
 $a,c\in s_1$ and $b,d\in s_2$,
 inconsistent in the binary patterns of their associated strings, {\em i.e.},
 $\eta_a+\eta_b\neq \eta_c+\eta_d$ with $\eta_a,\dots,\eta_d$ respectively denoting
 $a-1,\dots,d-1$ in their binary expressions.
 There should have one beginning row that contains
 the two integer pairs
 $(b,d)$ and $(a',c)$, leaving the integer $a'$ to be decided.
 They share the identical pattern $\eta_b+\eta_d=\eta_{a'}+\eta_c$,
 for these two integer pairs are in one beginning row and
 cling to the binary partitioning.
 The multiplication of these two rows of integer pairs produces,
 among the others, the pairs $(a,d)$ and $(a',d)$.
 Then, a contradiction occurs
 $\eta_{a'}=\eta_b+\eta_c+\eta_d\neq\eta_a$, which fails
 the condition of closure.
 The implication is hence derived that, to keep closed the multiplication of integer pairs,
 the interaction row must be obedient to the binary partition if
 the beginning rows have been so.
 As leaned earlier,
 the binary partitioning is fully respected by $su(4)$, whose
 subscript table is invariant under all integer permutations.
 The recursive argument thus starts with 
 $su(8),\hspace{2pt}p=3$,
 and let its conjugate pairs form according to the binary partitioning.
 Immediately it asserts that, for the Lie algebra $su(2^p)$, only
 the conjugate partition 
 following the binary partitioning
 admits the structure of the intrinsic quotient algebra of rank zero ${\cal \{Q(C};2^p-1)\}$.

 Even so, this assertion does not deny a {\em global} permutation that
 applies the same permutation of integers to every row of the table.
 For in a such circumstance still the binary partitioning is fulfilled
 and integers are permuted simply as labelling indices.
 It concludes the sufficient and necessary criterion that, up to a permutation
 on the subscripts of the $\lambda$-generators, 
 the conjugate partition admitting the quotient algebra
 ${\cal \{Q(C)\}}$ in $su(2^p)$ is unique,
 the one 
 conforming to the binary partitioning.
 That is to say, the complete set of conjugate partitions
 carrying ${\cal \{Q(C)\}}$ 
 is obtained by exhaustively permuting the subscripts of
     $\lambda$-generators initiated by the binary partitioning.
 However, these variants are composed of generators in
 superpositions of noncommuting spinors and of no current interest.

 \paragraph{Algebra Isomorphism.}
 As the dimension $N$ is not a power of $2$, the Lie algebra $su(N)$
 admits as well the structures of conjugate partitions
 and rank-zero quotient algebras. The reason can be read from 
 an example of a conjugate partition for $su(6)$ as in Fig.~\ref{Gsu6}.
 Written in the $\lambda$-representation similarly, this partition
 comprises $7$ conjugate pairs of abelian subspaces in addition to
 a center subalgebra of $5$ generators.
 An explicit distinction between Figs.~\ref{Gsu8} and~\ref{Gsu6} lies in the change that all the
 generators 
 with subscripts larger than the dimension $N=6$,
 {\em i.e.}, with $7$ or $8$ here, are removed from the latter.
 As suggested earlier, the generating set $\{d_{1l}:l=2,3,\dots,8\}$ is a convenient choice
 for the intrinsic center subalgebra of $su(8)$. 
 It has the interpretation that the $(i,j)$-th entry of an generator
 is responsible for contributing to transformations on the plane spanned over
 the $i$-th and the $j$-th dimensions.
 There are no the $7$th and the $8$th dimensions in $su(6)$ and transformations involving
 these two dimensions are forbidden. A generator with either one of these
 two subscripts then ought to be set to $0$ or simply removed from the partition.
 As a result, all the rest of $8\times8$ matrices of generators originally for $su(8)$
 are nil 
 in the $7$th and the $8$th rows and columns.
 Undergoing the removal of the $7$th and the $8$th rows and columns,
 a number $35$ of remaining generators of $6\times6$ matrices suffice for
 the formation of the algebra $su(6)$. 
 Conversely speaking, with the recoverage of the $7$th and the $8$th rows and columns
 despite being nil,
 these generators support an embedding of $su(6)$ in the space of $su(8)$.

 On the other hand, since only those with subscripts greater than
 the dimension $N=6$ are removed,
 all the $\lambda$-generators of subscripts less than or
 equal to the dimension survive the removing.
 Belonging to an abelian subspace of the intrinsic quotient algebra of $su(8)$ respectively,
 either in a conjugate pair or in the center subalgebra,
 these survived generators sustain the original conjugate-partition structure
 and realize the condition of closure all the same.
 An isomorphism  of the quotient algebra
 is therefore guaranteed for $su(6)$ and $su(8)$.
 This 
 {\bf \em removing process} is applicable to
 every Lie algebra $su(N)$ for $2^{p-1}<N<2^p$: Let the $\lambda$-generators
 $\lambda_{ij}$, $\hat{\lambda}_{ij}\in su(2^p)$ and those
 in ${\cal C}=Span\{d_{1l},l=2,3,\dots,2^p\}\subset su(2^p)$ be removed from
 the intrinsic quotient algebra $\{{\cal Q(C};2^p-1)\}$ of $su(2^p)$ iff
 $l, i\text{ or }j>N$.
 As a consequence, the Lie algebra $su(N)$ is spanned by a number $N^2-1$ of
 the remaining generators; from the $(N+1)$-th to
 the $2^p$-th rows and columns in the matrices of these generators are nil and allow a further removal.
 Still all the generators with subscripts ranging from $1$ to $N$
 reside in the conjugate pairs and the center subalgebra respectively,
 which are structurally inherited from
 the quotient algebra $\{{\cal Q(C};2^p-1)\}$. 
 The relations amongst these abelian subspaces of removed versions, Eqs.~\ref{D1}-\ref{D15},
 are preserved accordingly.
 Also, it is plain 
 to verify that none of the subspaces of $\{{\cal Q(C};2^p-1)\}$
 reduce to null due to the removing.
 The following corollary is hence asserted.
 \end{proof}
 \vspace{6pt}

  Unitary Lie algebras of certain dimensions enjoy the isomorphism of quotient algebras.
  \vspace{6pt}
  \begin{cor}\label{coro}
  The quotient algebra of rank zero of the Lie algebra $su(N)$, $2^{p-1}<N<2^p$,
  is isomorphic to that of $su(2^p)$. 
  \end{cor}
  \vspace{6pt}
  The concept of the binary partition is essential to characterizing the structure of
  a quotient algebra and is heavily engaged in the argument above, which however
  will become natural as being translated into
  the $s$-representation~\cite{SuTsai1}.
  Appendix~\ref{appsrep} may be consulted for a quick glance. 
  Not only performed as a basic tool in those of rank zero, also the removing process is
  applicable to quotient algebras of higher ranks.
  The extension leading to similar isomorphisms in higher ranks
  will be elaborated in the proof of Lemma~19 preceding to Main Theorem in~\cite{SuTsai2}.

\section{Scheme Implementation}\label{schmimpl}
\renewcommand{\theequation}{\arabic{section}.\arabic{equation}}
\setcounter{equation}{0} \noindent
  With the validation of Theorem~\ref{thm01}, the construction of a quotient algebra
  is complied into an algorithm as follows.

 \vspace{-.2cm}
 \paragraph{Algorithm.}
  {\bf Quotient Algebra $\{{\cal Q(A)}\}$ Construction} for $su(N)$, $2^{p-1}<N\leq 2^p$,\\
  $\hspace*{5.17cm}$with a center subalgebra ${\cal A}\subset su(N)$\\ 
  {\em The step of preparation:\\
  Find the unitary transformation $U$ that has ${\cal A}$ diagonalized,
  i.e., mapping ${\cal A}$ to the intrinsic center subalgebra ${\cal C}$
  by $U{\cal A}U^{\dagger}={\cal C}$.\\
  The 1st step{\rm :}\\
  Choose a generator $g_{11}\notin{\cal C}$ as a seed and calculate the
  commutator $[g_{11},{\cal C}]$, which produces a set of a finite number $\hat{r}_1$
  of abelian generators $\hat{{\bf g}}_1=
  \{\hat{g}_{11},\hat{g}_{12},\cdots,\hat{g}_{1\hat{r}_1}\}$, and then let the subspace
  $\hat{W}_1=<\hat{{\bf g}}_1>$ be spanned by $\hat{{\bf g}}_1$; 
  in the reverse step,
  the commutator $[\hat{g}_{1i},{\cal C}]$, $\hat{g}_{1i}\in\hat{{\bf g}}_1$,
  adds the other $r_1-1$ abelian generators to the set
  ${{\bf g}}_1=\{g_{11},g_{12},\cdots,g_{1r_1}\}$, which forms
  the conjugate subspace $W_1=<{{\bf g}}_1>$. \\
  .........\\
  The $l$-th step{\rm :} \\
  With a generator $g_{l1}\in su(N)-{\cal C}\bigcup^{l-1}_1\{W_i,\hat{W}_i\}$ taken
  as the $l$-th seed,
  the commutator $[g_{l1},{\cal C}]$ yields a set of abelian generators
  $\hat{{\bf g}}_l=\{\hat{g}_{l1},\hat{g}_{l2},\cdots,\hat{g}_{l\hat{r}_l}\}$
  and let the subspace
  $\hat{W}_l=<\hat{{\bf g}}_l>$ be spanned by the set $\hat{{\bf g}}_l$; in the reverse step,
  the commutator $[\hat{g}_{li},{\cal C}]$, $\hat{g}_{li}\in\hat{{\bf g}}_l$,
  contributes another abelian set 
  ${{\bf g}}_l=\{g_{l1},g_{l2},\cdots,g_{lr_l}\}$, which forms the conjugate subspace $W_l=<{{\bf g}}_l>$;\\
  Divide or merge the conjugate pair(s) into refined pair(s) according to the condition of closure whenever necessary.\\
  .........\\
  End when no generator remains.\\
  The construction of the intrinsic quotient algebra $\{{\cal Q(C)}\}$ 
  completes;\\
  Map $\{{\cal Q(C)}\}$ to the quotient algebra $\{{\cal Q(A)}\}$ 
  by the transformation $\{{\cal Q(A)}\}=U^{\dagger} \{{\cal Q(C)}\} U$.}
  \vspace{.5cm}


  The step of merging commuting conjugate pairs is necessary on occasion when
  a conjugate pair is created in part
  through calculating the commutator of the center subalgebra and a seed generator.
  Simple examples are available in constructing quotient algebras of $su(4)$
  in the $\lambda$-representation.
  By Eqs.~\ref{A1} and~\ref{A2}, the commutator $[g,{\cal C}]$ yields only
  a single generator $\hat{g}$, with $g$ being
  a $\lambda_{ij}$ or a $\hat{\lambda}_{ij}$ and ${\cal C}$ being the intrinsic
  center subalgebra of $su(4)$.
  There are two options of merging, for instance, either
  $\{\lambda_{12},\lambda_{34}\}\cup\{\hat{\lambda}_{12},\hat{\lambda}_{34}\}$
  or $\{\lambda_{12},\hat{\lambda}_{34}\}\cup\{\hat{\lambda}_{12},\lambda_{34}\}$
  for the two commuting conjugate pairs
  $\{\lambda_{12}\}\cup\{\hat{\lambda}_{12}\}$ and $\{\lambda_{34}\}\cup\{\hat{\lambda}_{34}\}$.
  Hence, in addition to the original one as in Figs.~\ref{csu4} and~\ref{Gsu4},
  the $2$nd version
  of the quotient algebra is rendered as in Fig.~\ref{AGsu4}. This is simply
  the {\em freedom within a conjugate pair} which allows the two distributions
  either $\lambda_{ij}\in W_l$ and $\hat{\lambda}_{ij}\in\hat{W}_l$ or
         $\lambda_{ij}\in\hat{W}_l$ and $\hat{\lambda}_{ij}\in{W}_l$ for
  each pair of generators $\{\lambda_{ij},\hat{\lambda}_{ij}\}$
  as long as conforming to the condition of closure. 
  However, the generators in the $2$nd version involve superpositions
  of noncommuting spinor generators, such as
  the part $(\sigma_1\pm \sigma_2)$ of the $2$nd qubit in spinor generators of a subspace
  $\lambda_{12}\pm \hat{\lambda}_{34}=\frac{1}{2}(I-\sigma_3)\otimes (\sigma_1\pm \sigma_2)$.
  A form of this kind is beyond the current interest.

  On the other hand, dividing existing conjugate pairs into refined pairs will become
  compelling while constructing (co-)quotient algebras of higher ranks~\cite{SuTsai2,SuTsai3}.
  Since center subalgebras in these occasions are a subset of a Cartan subalgebra,
  every subspace of a such (co-)quotient algebra is a refined subset partitioned from
  that of a quotient algebra of rank zero following the condition of closure Eq.~\ref{D1}.
  More often, the two subspaces of a conjugate pair are of the same size,
  {\em i.e.}, spanned by the same number of generators and thus $r_l=\hat{r}_l$. 
  In fewer cases of (co-)quotient algebras of higher ranks,
  one of the two subspaces of a conjugate pair may reduce to null.
  In practice, there is no order to generate conjugate pairs.
  Yet, it is not only essential but also helpful during the construction
  to check the commutation relations
  among the produced subspaces by the condition of closure.
  At certain stages for instance, 
  the seed generators needed are better learned by applying
  this condition to existing pairs.

  For the Lie algebra $su(2^p)$, a quotient algebra of rank zero 
  can be directly constructed with the algorithm by taking
  a non-intrinsic Cartan subalgebra 
  as the center subalgebra,
  and every the subspace $W_l$ or $\hat{W}_l$ is spanned by
  the same number $r_l=\hat{r}_l=2^{p-1}$ of generators.
  Starting with the intrinsic quotient algebra 
  and exercising the conjugate transformation is additionally required for that
  of dimension not being a power of $2$.
  Thanks to the algebra isomorphism and the removing process, there is
  an alternative way to build quotient algebras of rank zero $\{{\cal Q(A)}\}$
  for $su(N)$ as $2^{p-1}<N<2^p$.
  Compared with the former edition, this alternative is likely to be more efficient
  in particular when the dimension $N$ is large.
  Implemented on those of rank zero as an example, the procedure is as simple as follows.
  The first step is to derive the intrinsic quotient algebra $\{{\cal Q(C)}\}$
  of $su(N)$ from that $\{{\cal Q(\hat{C})}\}$ of $su(2^p)$ via conducting
  the removing process upon the latter, here ${\cal C}$ and ${\cal \hat{C}}$
  being the intrinsic Cartan subalgebras of $su(N)$ and $su(2^p)$ respectively.
  Then in the $2$nd step, the quotient algebra $\{{\cal Q(A)}\}$ is obtained
  right after the conjugate
  transformation $\{{\cal Q(A)}\}=U^{\dagger} \{{\cal Q(C)}\} U$, noticing that
  the Cartan subalgebras ${\cal A}\subset su(N)$ and ${\cal \hat{A}}\subset su(2^p)$
  share the identical conjugate action $U$ such that $U{\cal A}U^{\dagger}={\cal C}$ and
  $U{\cal \hat{A}}U^{\dagger}={\cal \hat{C}}$ with ${\cal A}\subset{\cal \hat{A}}$.
  Being the unique superset of the designated counterpart ${\cal A}\subset su(N)$,
  the Cartan subalgebra ${\cal \hat{A}}\subset su(2^p)$ can be easily 
  decided in terms of the $\lambda$-representation.
  As to the conjugate transformation,
  it is convenient to carry out the calculation in the
  $s$-representation, {\em ref}. Appendix~\ref{appsrep}, and the form
  of the action $U$ will be explicitly given in~\cite{SuTsai1,SuTsai3}.
  The procedure closes at rewriting $\{{\cal Q(A)}\}$ in the preferred representation.

 \paragraph{Cartan Decomposition.}
 One of immediate applications of (co-)quotient algebras is the Cartan decomposition~\cite{Helgason,Knapp}.
 The algebraic structure of a (co-)quotient algebra enables a systematic
 production of Cartan decompositions.
 A Cartan decomposition
 of the Lie algebra $su(N)$ is referred to
 as a decomposition $su(N)=\mathfrak{t}\oplus \mathfrak{p}$, where
 the subalgebra $\mathfrak{t}$ and the subset $\mathfrak{p}$ satisfy the condition
 $[\mathfrak{t},\mathfrak{t}]\subset \mathfrak{t}$,
 $[\mathfrak{t},\mathfrak{p}]\subset \mathfrak{p}$,
 $[\mathfrak{p},\mathfrak{p}]\subset \mathfrak{t}$
 and ${\rm Tr}\{\mathfrak{t}\hspace{.05cm}\mathfrak{p}\}=0$.
 The decomposition is not unique.
 Within a quotient algebra of rank zero of $2^p-1$ conjugate pairs,
 there have in total $2^p$ choices of
 Cartan decompositions.

  A Cartan decomposition chosen in a quotient algebra is equivalent to a determination of
  the subalgebra $\mathfrak{t}$, which is formed by taking the union of one
  subspace of every conjugate pair following the condition of closure.
  An example can be prepared in a rank-zero quotient algebra of either $su(8)$ or $su(6)$,
  which has $7$ conjugate pairs as shown in Figs.~\ref{csu8}-\ref{ncsu6}.
  The subalgebra $\mathfrak{t}$ is formed by taking either one subspace of
  every conjugate pair therein.
  However, due to the condition of closure, when the
  first two subspaces are chosen from the first two conjugate pairs,
  the  $3$rd one from the $3$rd pair is also decided.
  A reminder is that the conjugate pairs are orderless and the order assignment here
  is temporarily for the pair selection. Namely, any two conjugate pairs
  can be specified as the first two pairs, by which the $3$rd pair is decided due to
  the condition of closure.
  Likewise, let the $4$th conjugate pair be arbitrarily picked
  from the remaining pairs, which together with the early selections decides
  the following $5$th, $6$th and $7$th pairs.
  Then the subspaces from the $5$th, $6$th and $7$th conjugate pairs are determined
  once the choice from the $4$th conjugate pair is made.
  For instance, if the set $\{W_1,W_2,\hat{W}_3,\hat{W}_4\}$ is selected,
  the set to follow is only $\{W_5,W_6,\hat{W}_7\}$.
  The subalgebra $\mathfrak{t}$ is formed by the union of these two sets
  $\mathfrak{t} = \{W_1,W_2,\hat{W}_3,\hat{W}_4,W_5,W_6,\hat{W}_7\}$.
  The subset $\mathfrak{p}$ is the
  union of the center subalgebra ${\cal A}$ and the remaining subspace
  in every conjugate pair.
  It is plain to confirm the orthogonality ${\rm Tr}\{\mathfrak{t}\hspace{.05cm}\mathfrak{p}\}=0$
  of the subalgebra and the subset so arranged.
  There are in total $2^3$ choices of $\mathfrak{t}$
  from this given quotient algebra.
  In general, once more a demonstration of the {\bf \em pre-decision rule}
  on a quotient algebra of $2^p-1$ conjugate pairs,
  the other $2^p-p-1$ subspaces in the subalgebra $\mathfrak{t}$
  are determined by a number $p$ of {\em independent} selections
  following the condition of clousre Eq.~\ref{D1}.

  It is noticed that, besides the other $7$ alike, the center subalgebra ${\cal A}$
  is a maximal abelian subalgebra of $\mathfrak{p}$. This indicates the category
  that these decompositions are of type {\bf AI}~\cite{Helgason},
  because the maximal abelian subalgebra of
  the subset $\mathfrak{p}$ is a Cartan subalgebra of the original Lie algebra.
  It will be asserted in a continued episode~\cite{SuTsai2} that the complete set
  of type-{\bf AI} decompositions of a unitary Lie algebra
  can be acquired by collecting those determined from quotient algebras of rank zero
  generated by all its Cartan subalgebras.
  The determinations of the other two categoris of decompositions,
  the types {\bf AII} and {\bf AIII}, with quotient algebras of higher ranks
  will also be illustrated in the same episode.


\section{Recursive Decomposition}
\renewcommand{\theequation}{\arabic{section}.\arabic{equation}}
\setcounter{equation}{0} \noindent
  As a merit of the structure, a quotient algebra admits
  the recursive decomposition level by level.
  The formation of the above subalgebra $\mathfrak{t}$ is
  considered the {\em 1st-level} decomposition
  of $su(N)$ and let it be redenoted as $\mathfrak{t}_{[1]}$,
  noting $su(N)=\mathfrak{t}_{[0]}\oplus\mathfrak{p}_{[0]}$ the {\em 0th-level} decomposition of itself
  with $\mathfrak{t}_{[0]}=su(N)$ and $\mathfrak{p}_{[0]}=\{0\}$.
  Being a subset of abelian subspaces respecting the condition of closure equally well,
  the subalgebra $\mathfrak{t}_{[1]}$ bears the structure of the quotient algebra
  $\{{\cal Q(A}_{[2]};2^{p-1}-1)\}$, where ${\cal A}_{[2]}$ denotes
  an arbitrary maximal abelian subalgebra of $\mathfrak{t}_{[1]}$
  assigned as the center subalgebra of the $2$nd level.
  There are in total $2^{p-1}$ choices of the subalgebra of the $2$nd level
  $\mathfrak{t}_{[2]}$, for $\mathfrak{t}_{[1]}$
  is partitioned by ${\cal A}_{[2]}$ into $2^{p-1}-1$ conjugate pairs.
  Similarly, the quotient algebra $\{{\cal Q(A}_{[l]};2^{p-l}-1)\}=\mathfrak{t}_{[l-1]}$
  is constructed at the $l$-th-level decomposition for the subalgebra of the
  $(l-1)$-th level $\mathfrak{t}_{[l-1]}$, where
  the center subalgebra ${\cal A}_{[l]}$ is a maximal abelian subalgebra
  of $\mathfrak{t}_{[l-1]}$ and there has $2^{p-l}$ choices of the subalgebra
  of next level $\mathfrak{t}_{[l]}$.
  As this process is recursively
  continued for $p$ times to the final level and the quotient algebra
  $\{{\cal Q(A}_{[p]};1)\}$ is attained,
  the Lie algebra $su(N)$ is fully decomposed.

  Directing a such recursive decomposition, a {\bf\em decomposition sequence} $seq_{dec}$
  is defined to be the sequence of designated center subalgebras in the recursion order
  and at last the subalgebra of the final level, {\em i.e.},
  $seq_{dec}=\{{\cal A}_{[l]}:\hspace{.00cm}l=1,2,\cdots,p+1,\hspace{.06cm} {\cal A}_{[1]}={\cal A}\text{ and }
  {\cal A}_{[p+1]}=\mathfrak{t}_{[p]}\}$.
  Precisely speaking, in a decomposition sequence, the designated center algebra
  of the $l$-th level ${\cal A}_{[l]}$ ought to be a maximal abelian subalgebra of
  the subset $\mathfrak{p}_{[l]}$ of the decomposition of this level
  $\mathfrak{t}_{[l-1]}=\mathfrak{t}_{[l]}\oplus \mathfrak{p}_{[l]}$.
  In a decomposition of type {\bf AI}, the category concerned in this episode,
  a maximal abelian subalgebra of $\mathfrak{p}_{[l]}$ is also
  a maximal abelian subalgebra of $\mathfrak{t}_{[l-1]}$, $0<l\leq p$, and
  it is favorable to take
  an abelian subspace $W_i$ or $\hat{W}_i$
  of the original quotient algebra $\{{\cal Q(A)};2^{p}-1)\}$ to be the center subalgebra
  at every level~\cite{SuTsai1}.
  More complicated sequences involving different types of decompositions
  will be studied in continued episodes~\cite{SuTsai3}.


 \paragraph{Operator Factorization.}
  According to the $KAK$ theorem~\cite{Helgason,Knapp},
  a Cartan decomposition of the Lie algebra $su(N)$ leads to factorizations
  of actions in $SU(N)$.
  That is, there exist vectors ${\bf s_1},{\bf s_2}\in\mathfrak{t}$
  and ${\bf a}\in{\cal A}$ 
  for every ${\bf g}\in su(N)$ such that the corresponding group action
  $e^{i\mathfrak{g}}$ has a particular form of factorization
  $e^{i{\bf g}}=e^{i{\bf s_1}} e^{i{\bf a}} e^{i{\bf s_2}}$,
  if a Cartan decomposition $su(N)=\mathfrak{t}\oplus\mathfrak{p}$
  is realized 
  and ${\cal A}$ is a maximal abelian subalgebra of $\mathfrak{p}$.
  When the decomposition of type {\bf AI} is concerned,
  the computation of a such factorization is operationally
  equivalent to the well-known SVD (Singular Value Decomposition)
  in matrix analysis.

  As a consequence of repeating the $KAK$ theorem at each level,
  a recursive decomposition of the algebra $su(N)$, $2^{p-1}<N\leq 2^p$, 
  results in a recursive
  factorization of an action in $SU(N)$. 
  Through the factorization prescribed in the theorem,
  the action turns into a product of a number $2^{2p}-1$
  of component terms $e^{i{\bf a}_{[l],j}}$ contributed
  from the maximal abelian subalgebras of decomposition levels and
  these terms are arranged in the order of a {\em binary bifurcation tree}.
  A term of the form $e^{i{\bf a}_{[l],j}}$,
  ${\bf a}_{[l],j}\in{\cal A}_{[l]}$ as $1\leq l\leq p$ and ${\bf a}_{[p],j}\in{\bf t}_{[p]}$,
  is contributed from the center subalgebra designated at the $l$-th level or
  from the subalgebra of the final level.
  Let the order of the term in the factorization be written in
  a binary string of $p+1$ digits and
  denoted by $\tau_{lj}$, $0\leq l\leq p$ and $j=1,2,\dots,2^l$.
  The binary-bifurcation-tree order has the
  first bit ``1" of the string $\tau_{lj}$, counting from the rightmost,
  appear at the $(p-l+1)$-th digit.
  Some calculation facilities of factorizations led by decomposition sequences of type {\bf AI}
  will be given in~\cite{SuTsai1} and others by sequences of different types will be
  expounded in~\cite{SuTsai2,SuTsai3}.



  Following a decomposition sequence, the recursive decomposition of
  a transformation in $SU(N)$
  implies a {\em path} on the manifold of the group.
  A subalgebra designated in the sequence may be conceived as
  a {\em post} or a {\em node} during the course of the mapping.
  Rather than in a serial manner, these posts are visited in a
  {\em hierarchically recursive order}, patterned in a binary bifurcation tree,
  which exhibits the nature of recursive decompositions of Lie groups. 
  The course of a path on the manifold is adjustable by changing designations of subalgebras in the
  decomposition sequence. 
  To have a better picture of possible paths upon the group $SU(N)$,
  it is necessary to locate all the Cartan subalgebras of $su(N)$.
  The algorithm of the {\em subalgebra extension} introduced in Appendix~\ref{appabelext}
  suffices the need.

  Since every subalgebra in a decomposition sequence is abelian, a transformation
  in $SU(N)$ is fully factorized into a product of actions $e^{i\omega_\alpha g_\alpha}$.
  Wherein a $g_\alpha$ is a generator in the chosen representation and $\omega_\alpha$
  is the parameter acquired by such as the operation of SVD
  if the type-{\bf AI} decomposition is concerned.
  Suppose that the dimension $N$ has the prime factorization
  $N=2^{r_0}{P_1}^{r_1}\cdots {P_f}^{r_f}$, here
  $P_j$, $j=1,\dots,f$, being a prime greater than $2$.
  An option of convenience is to put the generators of $su(N)$ in tensor products of
  the spinor generators and those of $su(P_j)$ phrased in
  the $\lambda$-representation, {\em ref}. Appendix A.
  Each algebra $g_\alpha$ thus reads as
  a tensor product of Pauli martices, $\lambda$-generators and
  an appropriate number of identity operators.

  Playing a role of a quantum gate,
  a contributed action $e^{i\omega_\alpha g_\alpha}$ is considered {\em local} if it has only one single
  non-identity operator, {\em i.e.}, a generator of $su(2)$ or $su(P_j)$, in the
  tensor-product form 
  of $g_\alpha$, and considered {\em nonlocal} if otherwise.
  Accordingly, a decomposition of the associated algebra as above described makes
  an arbitrary unitary transformation factorized into a product of
  local and nonlocal transformations.
  Moreover, it is straightforward to further factorize a nonlocal transformation into
  a product of local and bipartite actions of an any specified format 
  resorting to some elementary identities of Lie algebras~\cite{SuTsai3}; 
  a bipartite action here is referring to a transofmration involving two parties such as,
  but not restricted to, the {CNOT} gate.
  As an immediate result of an application to an arbitrary unitary operator,
  this decomposition scheme also provides 
  an alternative and simple proof for the
  computational universality of quantum gates.
  Refer to~\cite{SuTsai2} for a general proof of the universality.
  \vspace{6pt}
  \begin{cor}\label{coro2}
  An any bipartite gate plus one qubit is computaionally universal.
  \end{cor}
  \vspace{6pt}
  Since 
  being flexible and offering great freedom to decide paths of decompositions,
  structures of quotient algebras expounded 
  in this serial are anticipated to be of help to
  many inquiries, not merely confined to determining optimal gate arrangements subject to certain
  imposed constraints.
  More details of various applications derived from these structures and related schemes
  will be systematically elucidated elsewhere~\cite{SuTsai1,SuTsai2,SuTsai3,SuTsaiQAPFT,SuTsaiOptFTQC}.

\nonumsection{References}

 \appendix{~~$\lambda$-Representation\label{applambdarep}}
 A $\lambda$-generator,
 $\lambda_{ij}$ or its conjugate $\hat{\lambda}_{ij}$,
 of the $\lambda$-representation is an off-diagonal $N\times N$ matrix
 and serves the role of $\sigma_1$ or $\sigma_2\in su(2)$ in
 the $i$-th and the $j$-th dimensions.
 In terms of the Dirac notation, the generators read as
 $\lambda_{ij}=\ket{i}\bra{j}+\ket{j}\bra{i}$ and
 $\hat{\lambda}_{ij}=-i\ket{i}\bra{j}+i\ket{j}\bra{i}$.
 Playing the part of $\sigma_3$, a diagonal $N\times N$ matrix $d_{kl}$ is
 defined as $d_{kl}=\ket{k}\bra{k}-\ket{l}\bra{l}$.
 A generating set of the Lie algebra $su(N)$ can
 be formed by the $N(N-1)/2$ pairs of
 $\lambda_{ij}$ and $\hat{\lambda}_{ij}$ in addition to
 any $N-1$ independent diagonal operators $d_{kl}$,
 such as $d_{1l}, l=2,3,\dots,N$.
 It is plain to confirm that
 ${\rm Tr}\{\breve{\lambda}_{ij}\breve{\lambda}_{kl}\}=0$ and
 ${\rm Tr}\{\breve{\lambda}_{ij}{\cal C}\}=0$,
 $\breve{\lambda}_{ij}\neq\breve{\lambda}_{kl}$,
 where $\breve{\lambda}_{ij}$ denotes either
 a $\lambda_{ij}$ or a $\hat{\lambda}_{ij}$
 and ${\cal C}$ is the intrinsic center subalgebra of $su(N)$.
 Since ${\cal C}\subset\mathfrak{p}$, the condition of orthogonality ${\rm Tr}\{\mathfrak{t} \mathfrak{p}\}=0$
 is guaranteed at each level of the decomposition.
 Although not compulsory in the decomposition,
 the orthogonality can be
 further established in ${\cal C}$ by taking a certain set of basis generators,
 {\em e.g.},
 ${\cal C}=span\{\sqrt{\frac{2}{l(l-1)}}\sum_{i=1}^{l-1} d_{il};l=2,3,\dots,N\}$.

 The $\lambda$-representation
 is in particular a suitable choice when the dimension $N$ is a prime.
 An example is the set of the Gell-Mann matrices for $su(3)$,
 here denoted by $\mu_{j}$, $1\leq j\leq 8$,
\begin{align*}
\mu_{1}=\lambda_{12} \text{,}& \hspace{0.5cm}\mu_{2}=\hat{\lambda}_{12},\\
\mu_{4}=\lambda_{13} \text{,}& \hspace{0.5cm}\mu_{5}=\hat{\lambda}_{13},\\
\mu_{6}=\lambda_{23} \text{,}& \hspace{0.5cm}\mu_{7}=\hat{\lambda}_{23},\\
\mu_{3}=d_{12}       \text{,}&
\hspace{0.5cm}\mu_{8}=\frac{1}{\sqrt{3}}(d_{13}+d_{23}).
\end{align*}
 This representation has advantages as well
 even if the dimension $N$ is a composite number, say
 $N=2^{r_0}{P_1}^{r_1}\cdots {P_f}^{r_f}$
  and $P_j$ being primes greater than $2$, $j=1,\dots, f$.
 First, let the table of a quotient algebra for $su(2^{r_0})$ be produced
 by taking an arbitrary center subalgebra in the spinor representation and
 exercising the algorithm instructed in Section~\ref{schmimpl}.
 Before starting the algorithm, a quick way to build a similar table for
 $su(N)$ is then to write the center subalgebra in tensor products
 of the spinor generators and the generators of those $su(P_j)$ in
 the $\lambda$-representation.

 The following commutation identities characterize the $\lambda$-generators and
 are crucial to the construction of conjugate partitions and quotient algebras,
 $1\leq i,j,k,l\leq N$,
\begin{align}
 [\lambda_{ij},d_{kl}]=&\hspace{0.5cm} i\hat{\lambda}_{ij} (-\delta_{ik}+\delta_{il}+\delta_{jk}-\delta_{jl})\label{A1}\\
 [\hat{\lambda}_{ij},d_{kl}]=&\hspace{0.5cm} i\lambda_{ij} (\ \ \delta_{ik}-\delta_{il}-\delta_{jk}+\delta_{jl})\label{A2}\\
 [\lambda_{ij},\hat{\lambda}_{ij}]=&\hspace{0.5cm} 2i d_{ij}\label{A3}\\
[\lambda_{ij},\lambda_{kl}]=&\hspace{0.5cm} i\hat{\lambda}_{ik}\delta_{jl}+i\hat{\lambda}_{il}\delta_{jk}+i\hat{\lambda}_{jk}\delta_{il}+i\hat{\lambda}_{jl}\delta_{ik}\label{A4}\\
[\lambda_{ij},\hat{\lambda}_{kl}]=&\hspace{0.5cm} i\lambda_{ik}\delta_{jl}-i\lambda_{il}\delta_{jk}+i\lambda_{jk}\delta_{il}-i\lambda_{jl}\delta_{ik}\label{A5}\\
[\hat{\lambda}_{ij},\hat{\lambda}_{kl}]=&\hspace{0.5cm}
i\hat{\lambda}_{ik}\delta_{jl}-i\hat{\lambda}_{il}\delta_{jk}-i\hat{\lambda}_{jk}\delta_{il}+i\hat{\lambda}_{jl}\delta_{ik}.\label{A6}
\end{align}
 A $\lambda$-generator is either symmetric or antisymmetric
 with respect to the subscript interchange,
 {\em i.e.}, $\lambda_{ij}=\lambda_{ji}$,
 $\hat{\lambda}_{kl}=-\hat{\lambda}_{lk}$ and $d_{kl}=-d_{lk}$.
 This is for the completion of definition and only those members of
 the latter subscript greater than the former are in actual use.

 Required for later purposes~\cite{SuTsai2}, also the anti-commutation identities
 are recorded as follows,
 $1\leq i,j,k,l\leq N$,

\appendix{~~{\em s}-Representation}\label{appsrep}
 The {\em s-representation} is an expression of convenience for spinor generators.
 Let a $1$-qubit spinor be rephrased in digits, namely
 ${\cal S}^{\epsilon_1}_{a_1}=(\ket{0}\bra{a_1}+(-1)^{\epsilon_1}\ket{1}\bra{1+a_1})$,
 where the bit 
 $a_1$ or $\epsilon_1$ is either $0$ or $1$.
 Specifically, the elementary Pauli matrices are rewritten in the forms
 ${\cal S}^{0}_{\hspace{.81pt}0}=I$, ${\cal S}^{1}_{\hspace{.9pt}0}=\sigma_{3}$,
 ${\cal S}^{0}_{\hspace{.92pt}1}=\sigma_{1}$ and $-i{\cal S}^{1}_{\hspace{.92pt}1}=\sigma_{2}$.
 Being a tensor product of $1$-qubit shares, a $p$-qubit spinor generator is characterized by two $p$-digit binary strings
 $\alpha=a_1 a_2\ldots a_{p}$ and $\zeta=\epsilon_1\epsilon_2\ldots\epsilon_{p}$,
 \begin{align}\label{eqSrepinapp}
 {\cal S}^{\zeta}_{\alpha}={\cal S}^{\epsilon_{1}\epsilon_2\ldots
 \epsilon_{p}}_{a_{1} a_2\ldots a_{p}}
 ={\cal S}^{\epsilon_{1}}_{a_{1}}\otimes{\cal S}^{\epsilon_{2}}_{a_{2}}\ldots\otimes{\cal S}^{\epsilon_{p}}_{a_{p}}
 =\bigotimes^{p}_{i=1}\ (\ket{0}\bra{a_{i}}+(-1)^{\epsilon_{i}}\ket{1}\bra{1+a_{i}})\text{;}
 \end{align}
 for instances, ${\cal S}^{010}_{101}=\sigma_{1}\otimes\sigma_{3}\otimes\sigma_{1}$ and
 $-i{\cal S}^{0100}_{0111}=I\otimes\sigma_{2}\otimes\sigma_{1}\otimes\sigma_{1}$.
 The subscript string $\alpha$ indicates
 the {\em binary partitioning} to which the spinor generator is associated,
 and the superscript string $\zeta$ is
 interpreted as the {\em phase} of the generator.

 In the $s$-representation, the multiplication of two arbitrary spinors
 ${\cal S}^{\zeta}_{\alpha}$ and ${\cal S}^{\eta}_{\beta}$ exhibits the feature
 of the {\em bi-addition}~\cite{SuTsai1},
 \begin{align}\label{eqSproduct}
 {\cal S}^{\zeta}_{\alpha}\cdot{\cal S}^{\eta}_{\beta}
 =(-1)^{\eta\cdot\alpha}{\cal S}^{\zeta+\eta}_{\alpha+\beta}\text{.}
 \end{align}
 Note that the representation allows the {\em inner product} of two binary strings of the same length, {\em e.g.,}
 $\eta\cdot\alpha=\sum^{p}_{i=1}\sigma_{i} a_{i}$ and
 $\zeta\cdot\beta=\sum^{p}_{i=1}\epsilon_{i} b_{i}$, where
 $\beta=b_1 b_2 \ldots b_p$ and $\eta=\sigma_1\sigma_2\ldots\sigma_p$.
 Owing to the product form of Eq.~\ref{eqSrepinapp}, it suffices to affirm
 the relation of Eq.~\ref{eqSproduct} by verifying the multiplcation of
 two 1-qubit generators for $1\leq i\leq p$,
 \begin{align}\label{eqsmulti}
 &\hspace{-1pt}{\cal S}^{\epsilon_i}_{a_i}\cdot{\cal S}^{\sigma_i}_{b_i}\notag\\
 &\hspace{-5pt}=(\ket{0}\bra{a_i}+(-1)^{\epsilon_i}\ket{1}\bra{1+a_i})\cdot(\ket{0}\bra{b_i}+(-1)^{\sigma_i}\ket{1}\bra{1+b_i})\notag\\
 &\hspace{-5pt}=\delta_{0a_i}\ket{0}\bra{b_i}+\delta_{1a_i}(-1)^{\sigma_i}\ket{0}\bra{1+b_i}+\delta_{1a_i}(-1)^{\epsilon_i}\ket{1}\bra{b_i}+\delta_{0a_i}(-1)^{\epsilon_i+\sigma_i}\ket{1}\bra{1+b_i}\notag\\
 &\hspace{-5pt}=\delta_{0a_i}\ket{0}\bra{a_i+b_i}+\delta_{1a_i}(-1)^{\sigma_i}\ket{0}\bra{a_i+b_i}+\delta_{1a_i}(-1)^{\epsilon_i}\ket{1}\bra{1+a_i+b_i}\notag\\
 &\hspace{4pt}+\delta_{0a_i}(-1)^{\epsilon_i+\sigma_i}\ket{1}\bra{1+a_i+b_i}\notag\\
 &\hspace{-5pt}=(\delta_{0a_i}+(-1)^{\sigma_i}\delta_{1a_i})(\ket{0}\bra{a_i+b_i}+(-1)^{\epsilon_i+\sigma_i}\ket{1}\bra{1+a_i+b_i})\notag\\
 &\hspace{-5pt}=(-1)^{\sigma_i\cdot a_i}{\cal S}^{\epsilon_i+\sigma_i}_{a_i+b_i}.
 \end{align}
 An immediate implication is the commutativity of two spinors,
 \begin{align}\label{eqcommanticommspinors}
  {\cal S}^{\zeta}_{\alpha}{\cal S}^{\eta}_{\beta}
  =(-1)^{\eta\cdot\alpha+\zeta\cdot\beta}{\cal S}^{\eta}_{\beta}{\cal S}^{\zeta}_{\alpha}
  =
  \begin{cases}
      \hspace{9pt}{\cal S}^{\eta}_{\beta}{\cal S}^{\zeta}_{\alpha}    & \quad \text{if } \eta\cdot\alpha+\zeta\cdot\beta=0,2;\\
     -{\cal S}^{\eta}_{\beta}{\cal S}^{\zeta}_{\alpha}                & \quad \text{if } \eta\cdot\alpha+\zeta\cdot\beta=1,3.
  \end{cases}
 \end{align}
 There thus implies {\em the parity condition} to decide whether two spinors commute.
 That is, The {\em parity condition} reveals that
  two spinors {\em commute} if $\eta\cdot\alpha+\zeta\cdot\beta=0,2$
  or {\em anticommute} if $\eta\cdot\alpha+\zeta\cdot\beta=1,3$.
 These relations and condition will be fundamental to guiding constructions
 of quotient algebra partition in
 continued episodes~\cite{SuTsai1,SuTsai2,SuTsai3}.

  A spinor ${\cal S}^{\zeta}_{\alpha}\in{su(2^p)}$
  transforms a basis state into another
 \begin{align}\label{spinoronstate}
  {\cal S}^{\zeta}_{\alpha}\ket{\beta}
 =(-1)^{\zeta\cdot(\beta+\alpha)}\ket{\beta+\alpha}.
 \end{align}
 Likewise, this transformation is affirmed on the occasion of 1 qubit,
 {\em i.e.}, for $1\leq i\leq p$,
 \begin{align}\label{1Qubitspinoronstate}
 &{\cal S}^{\epsilon_i}_{a_i}\ket{b_i}\notag\\
 =&\hspace{1pt} ( \hspace{1pt} \ket{0}\bra{a_i}+(-1)^{\epsilon_i}\ket{1}\bra{1+a_i} \hspace{1pt})\hspace{2pt}\ket{b_i}\notag\\
 =&\hspace{2pt}\delta_{a_i,b_i}\ket{0}+(-1)^{\epsilon_i}\delta_{1+a_i,b_i}\ket{1}\notag\\
 =&\hspace{2pt}\delta_{0,b_i+a_i}\ket{0}+(-1)^{\epsilon_i}\delta_{1,b_i+a_i}\ket{1}\notag\\
 =&(-1)^{\epsilon_i\cdot (b_i+a_i)}\ket{b_i+a_i}.
 \end{align}
 Whereas a spinor ${\cal S}^{\zeta}_{\alpha}$
 is a generator in the algebra $su(2^p)$, but also a group action
 in $SU(2^p)$ thanks to the exponential mapping
 $i(-i)^{\zeta\cdot\alpha}{\cal S}^{\zeta}_{\alpha}=e^{i\frac{\pi}{2}(-i)^{\zeta\cdot\alpha}{\cal S}^{\zeta}_{\alpha}}$
 as in Eq.~\ref{s-rots}.
  A spinor $(-i)^{\zeta\cdot\alpha}{\cal S}^{\zeta}_{\alpha}$,
  multiplied by the phase $(-i)^{\zeta\cdot\alpha}$, is {\em hermitian},
  {\em i.e.},
  $((-i)^{\zeta\cdot\alpha}{\cal S}^{\zeta}_{\alpha})^{\dag}=(-i)^{\zeta\cdot\alpha}{\cal
  S}^{\zeta}_{\alpha}$.
  To retain the hermiticity,
 the inner product $\zeta\cdot\alpha$ counts the number of
 occurences of the 1-qubit component ${\cal S}^1_1$ in an
 $n$-qubit spinor
 ${\cal S}^{\zeta}_{\alpha}=\bigotimes^p_{i=1}{\cal S}^{\epsilon_i}_{a_i}$,
 and the accumulated exponent $\zeta\cdot\alpha$ of the phase
 $(\pm i)^{\zeta\cdot\alpha}$ is modulo $4$.
 While, this phase is often
 neglected if no confusion arises.
 Owing to Eq.~\ref{eqSproduct},  a spinor is {\em involutory}
 \begin{align}\label{eqInvotory}
 ((-i)^{\zeta\cdot\alpha}{\cal S}^{\hspace{.5pt}\zeta}_{\alpha})^2
 =(-1)^{\zeta\cdot\alpha}(-1)^{\zeta\cdot\alpha}{\cal S}^{\zeta+\zeta}_{\alpha+\alpha}
 ={\cal S}^{\mathbf{0}}_{\mathbf{0}}.
 \end{align}

   An {\em $s$-rotation} ${\cal R}^{\zeta}_{\alpha}(\theta)=e^{i\theta(-i)^{\zeta\cdot\alpha}{\cal S}^{\zeta}_{\alpha}}$
   of a spinor $(-i)^{\zeta\cdot\alpha}{\cal S}^{\zeta}_{\alpha}$
   has the expression
 \begin{align}\label{s-rots}
 e^{i\theta(-i)^{\zeta\cdot\alpha}{\cal S}^{\zeta}_{\alpha}}
 =cos\theta\hspace{1pt}{\cal S}^{\mathbf{0}}_{\mathbf{0}}+isin\theta\hspace{1pt}(-i)^{\zeta\cdot\alpha}{\cal S}^{\zeta}_{\alpha},
 \end{align}
 ${\cal S}^{\mathbf{0}}_{\mathbf{0}}\in{su(2^n)}$ being the identity and $0\leq\theta <2\pi$.
 The derivation is straightforward
 \begin{align}\label{ExpSpinor}
  &e^{i\theta(-i)^{\zeta\cdot\alpha}{\cal S}^{\zeta}_{\alpha}}\notag\\
 =&{\cal S}^{\mathbf{0}}_{\mathbf{0}}+i\theta(-i)^{\zeta\cdot\alpha}{\cal S}^{\zeta}_{\alpha}
 +\frac{1}{2!}(i\theta(-i)^{\zeta\cdot\alpha}{\cal S}^{\zeta}_{\alpha})^2
 +\frac{1}{3!}(i\theta(-i)^{\zeta\cdot\alpha}{\cal S}^{\zeta}_{\alpha})^3
 +\frac{1}{4!}(i\theta(-i)^{\zeta\cdot\alpha}{\cal S}^{\zeta}_{\alpha})^4
 +\cdots\notag\\
 =&{\cal S}^{\mathbf{0}}_{\mathbf{0}}+i\theta(-i)^{\zeta\cdot\alpha}{\cal S}^{\zeta}_{\alpha}
 -\frac{1}{2!}\theta^2{\cal S}^{\mathbf{0}}_{\mathbf{0}}
 -\frac{1}{3!}i\theta^3(-i)^{\zeta\cdot\alpha}{\cal S}^{\zeta}_{\alpha}
 +\frac{1}{4!}\theta^4{\cal S}^{\mathbf{0}}_{\mathbf{0}}
 +\cdots\notag\\
 =&\{ 1-\frac{1}{2!}\theta^2+\frac{1}{4!}\theta^4+\cdots  \}{\cal S}^{\mathbf{0}}_{\mathbf{0}}
 + i\{ \theta-\frac{1}{3!}\theta^3+\frac{1}{5!}\theta^5+\cdots  \}(-i)^{\zeta\cdot\alpha}{\cal S}^{\zeta}_{\alpha}
 \notag\\
 =&cos\theta{\cal S}^{\mathbf{0}}_{\mathbf{0}}+isin\theta(-i)^{\zeta\cdot\alpha}{\cal
 S}^{\zeta}_{\alpha}
 \end{align}
 invoking the involutory of a spinor
 as in Eq.~\ref{eqInvotory}.
 Remark that the $s$-rotation
 $e^{i\theta(-i)^{\zeta\cdot\alpha}{\cal S}^{\zeta}_{\alpha}}$
 is an exponential mapping of $(-i)^{\zeta\cdot\alpha}{\cal S}^{\zeta}_{\alpha}$,
 a rotation acting on the space of $p$-qubit states,
 albeit not a rotation about the axis along this spinor.
 By applying an $s$-rotation ${\cal R}^{\zeta}_{\alpha}(\theta)$
 to a spinor $(-i)^{\eta\cdot\beta}{\cal S}^{\eta}_{\beta}$,
 it obtains
\begin{align}\label{spintospinTrans}
  &{\cal R}^{\zeta\dag}_{\alpha}(\theta)\hspace{2pt}(-i)^{\eta\cdot\beta}{\cal S}^{\eta}_{\beta}\hspace{2pt}{\cal R}^{\zeta}_{\alpha}(\theta)\notag\\
 =&(-i)^{\eta\cdot\beta}\{ cos\theta\hspace{1pt}{\cal S}^{\mathbf{0}}_{\mathbf{0}}-isin\theta\hspace{1pt}(-i)^{\zeta\cdot\alpha}{\cal S}^{\zeta}_{\alpha} \}
 \hspace{2pt}{\cal S}^{\eta}_{\beta}\hspace{2pt}
 \{ cos\theta\hspace{1pt}{\cal S}^{\mathbf{0}}_{\mathbf{0}}+isin\theta\hspace{1pt}(-i)^{\zeta\cdot\alpha}{\cal S}^{\zeta}_{\alpha} \}
 \notag\\
 =&(-i)^{\eta\cdot\beta}\{ cos\theta\hspace{1pt}{\cal S}^{\eta}_{\beta}-isin\theta\hspace{1pt}(-i)^{\zeta\cdot\alpha}(-1)^{\eta\cdot\alpha}{\cal S}^{\zeta+\eta}_{\alpha+\beta} \}
 \hspace{2pt}
 \{ cos\theta\hspace{1pt}{\cal S}^{\mathbf{0}}_{\mathbf{0}}+isin\theta\hspace{1pt}(-i)^{\zeta\cdot\alpha}{\cal S}^{\zeta}_{\alpha} \}
 \notag\\
 =&\hspace{10pt}(-i)^{\eta\cdot\beta}\hspace{2pt}\{ cos^2\theta\hspace{1pt}{\cal S}^{\eta}_{\beta}+sin^2\theta\hspace{1pt}(-1)^{\eta\cdot\alpha+\zeta\cdot\alpha}
 (-1)^{\zeta\cdot(\alpha+\beta)}{\cal S}^{\eta}_{\beta}\notag\\
 &+icos\theta\hspace{1pt}sin\theta\hspace{1pt}
 (-i)^{\zeta\cdot\alpha}(-1)^{\zeta\cdot\beta}{\cal S}^{\zeta+\eta}_{\alpha+\beta}
 -isin\theta cos\theta\hspace{1pt}(-i)^{\zeta\cdot\alpha}(-1)^{\eta\cdot\alpha}{\cal S}^{\zeta+\eta}_{\alpha+\beta} \}
 \notag\\
 =&(-i)^{\eta\cdot\beta}\{ (cos^2\theta+(-1)^{\eta\cdot\alpha+\zeta\cdot\beta}sin^2\theta){\cal S}^{\eta}_{\beta}
 +\frac{i}{2}(-i)^{\zeta\cdot\alpha}sin2\theta\hspace{1pt}((-1)^{\zeta\cdot\beta}-(-1)^{\eta\cdot\alpha}){\cal S}^{\zeta+\eta}_{\alpha+\beta}\}  \notag\\
 =&(-i)^{\eta\cdot\beta}\{ (cos^2\theta+(-1)^{\eta\cdot\alpha+\zeta\cdot\beta}sin^2\theta){\cal S}^{\eta}_{\beta}
 +\frac{i}{2}(-i)^{\zeta\cdot\alpha}(-1)^{\zeta\cdot\beta}sin2\theta\hspace{1pt}(1-(-1)^{\eta\cdot\alpha+\zeta\cdot\beta}){\cal S}^{\zeta+\eta}_{\alpha+\beta}\}\notag\\
 =&\hspace{10pt}(cos^2\theta+(-1)^{\eta\cdot\alpha+\zeta\cdot\beta}sin^2\theta)\hspace{2pt}(-i)^{\eta\cdot\beta}{\cal S}^{\eta}_{\beta}\notag\\
  &+\frac{i}{2}(-1)^{\zeta\cdot\beta}(-i)^{\zeta\cdot\alpha+\eta\cdot\beta}
    sin2\theta\hspace{1pt}(1-(-1)^{\eta\cdot\alpha+\zeta\cdot\beta}){\cal  S}^{\zeta+\eta}_{\alpha+\beta}.
 \end{align}
 Thus, the operator remains invariant if $\eta\cdot\alpha+\zeta\cdot\beta=0,2$,
 namely
 \begin{align}\label{spintospinTransComm}
  &{\cal R}^{\zeta\dag}_{\alpha}(\theta)\hspace{2pt}(-i)^{\eta\cdot\beta}{\cal S}^{\eta}_{\beta}\hspace{2pt}{\cal R}^{\zeta}_{\alpha}(\theta)
 =(-i)^{\eta\cdot\beta}{\cal S}^{\eta}_{\beta}.
 \end{align}
  If $\eta\cdot\alpha+\zeta\cdot\beta=1,3$,
  the spinor is transformed into
 \begin{numcases}{({\cal R}^{\zeta}_{\alpha}(\theta))^{\dag}
 \hspace{2pt}(-i)^{\eta\cdot\beta}{\cal S}^{\eta}_{\beta}\hspace{2pt}
 {\cal R}^{\zeta}_{\alpha}(\theta)=}
  \hspace{46pt}-(-i)^{\eta\cdot\beta}{\cal S}^{\eta}_{\beta}
  & as  \hspace{6pt}$\theta=\pm\frac{\pi}{2}$\label{eqsRotpi/2};\\
  \pm \varrho\cdot(-i)^{(\zeta+\eta)\cdot(\alpha+\beta)}{\cal S}^{\zeta+\eta}_{\alpha+\beta}
  &  as $\hspace{6pt}\theta=\pm\frac{\pi}{4}$\label{eqsRotpi/4},
\end{numcases}
 the coefficient
 $\varrho=i(-1)^{\zeta\cdot\beta}(-i)^{\zeta\cdot\alpha+\eta\cdot\beta}(i)^{(\zeta+\eta)\cdot(\alpha+\beta)}=\pm  1$.
 Remind that the hermiticity
 is maintained by the arithmetic that
 the inner product $\zeta\cdot\alpha$ counts the number of
 the 1-qubit component ${\cal S}^1_1$ occurring in
 an $n$-qubit spinor ${\cal S}^{\zeta}_{\alpha}$,
 and the accumulated exponent $\zeta\cdot\alpha$
 of a phase $(\pm i)^{\zeta\cdot\alpha}$  is modulo $4$.

 For the Lie algebra $su(N)$ of dimension $2^{p-1}<N<2^p$ not being a power of $2$,
 all the corresponding conditioned subspaces in the quotient-algebra structure
 can be obtained by applying the removing process.
 In practical calculations, it requires also a reverse process, the {\em embedding},
 which has $su(N)$ be embedded in the space of $su(2^p)$.
 In an operational sense, a such embedding process is simply to rewrite the
 generators of $su(N)$ in terms of the $s$-representation.
 The set of $\lambda$-generators is a suited choice in general as a generating set for
 $su(N)$. To relate to the $s$-representation,
 the subscripts of the $\lambda$-generators are relabelled in $p$-digit strings
 such that
 ${\cal L}^{\hspace{1pt}0}_{\hspace{1pt}\omega,\alpha}=i\hat{\lambda}_{i,j}$,
 ${\cal L}^{\hspace{1pt}1}_{\hspace{1pt}\omega,\alpha}=\lambda_{i,j}$
 and ${\cal L}^{\hspace{1pt}1}_{\hspace{1pt}\omega,{\bf 0}}-
      {\cal L}^{\hspace{1pt}1}_{\hspace{1pt}\omega+\alpha,{\bf 0}}=2d_{i,j}$, where
 ${\cal L}^{\hspace{1pt}\epsilon}_{\hspace{1pt}\omega,\alpha}
 \equiv\ket{\hspace{1pt}\omega\hspace{1pt}}\bra{\hspace{1pt}\omega+\alpha\hspace{1pt}}
 +(-1)^{1+\epsilon}\ket{\hspace{1pt}\omega
 +\alpha\hspace{1pt}}\bra{\hspace{1pt}\omega\hspace{1pt}}$,
 $i-1=\omega$, $j-1=\omega+\alpha$, $\omega,\alpha\in Z^p_2$ and $\epsilon=0,1$.
 Apparently, the generator ${\cal L}^{\hspace{1pt}\epsilon}_{\hspace{1pt}\omega,\alpha}$
 belongs to the conditioned subspace $W_{\alpha}$ or $\hat{W}_{\alpha}$ in
 the quotient algebra $\{{\cal Q}(\mathfrak{C}_{[\mathbf{0}]})\}$
 of the intrinsic Cartan subalgebra $\mathfrak{C}_{[\mathbf{0}]}$.
 The embedding of $su(N)$, $2^{p-1}<N<2^p$, to $su(2^p)$ is 
 realized by the substitution via the following formulae,
 here $\zeta,\alpha,\omega\in Z^p_2$ and $\epsilon\in{Z_2}$,
 \begin{align}\label{eqembS1}
 {\cal S}^{\zeta}_{\alpha}
 =(1/2)\sum_{\omega\in{Z^p_2},\hspace{1pt}\epsilon=1+\zeta\cdot\alpha}(-1)^{\zeta\cdot\omega}{\cal L}^{\hspace{1pt}\epsilon}_{\hspace{1pt}\omega,\alpha}
 \end{align}
 and
 \begin{equation}\label{eqembS2}
 \hspace{-29pt}{\cal L}^{\hspace{1pt}\epsilon}_{\hspace{1pt}\omega,\alpha}
 =(1/2^{p-1})\sum_{\zeta\in{Z^p_2},\hspace{1pt}\zeta\cdot\alpha=1+\epsilon}(-1)^{\zeta\cdot\omega}{\cal S}^{\xi}_{\alpha}.
 \end{equation}

 The commutator and
 the anti-commutator of two $\lambda$-generators 
 for later purposes~\cite{SuTsai2} are prepared as follows,
 \begin{align}\label{eqcommL}
 &[{\cal L}^{\hspace{1pt}\epsilon}_{\hspace{1pt}\omega,\alpha},{\cal L}^{\hspace{1pt}\sigma}_{\hspace{1pt}\tau,\beta}]\notag\\
 &\hspace{-3pt}=(\delta_{\omega+\alpha,\tau}+(-1)^{\sigma}\delta_{\omega+\alpha,\tau+\beta}){\cal L}^{\hspace{1pt}\epsilon+\sigma}_{\hspace{1pt}\omega,\alpha+\beta}
 +(-1)^{\epsilon}(\delta_{\omega,\tau}+(-1)^{\sigma}\delta_{\omega,\tau+\beta}){\cal L}^{\hspace{1pt}\epsilon+\sigma}_{\hspace{1pt}\omega+\alpha,\alpha+\beta}
 \end{align}
 and
 \begin{align}\label{eqanticommL}
 &\{{\cal L}^{\hspace{1pt}\epsilon}_{\hspace{1pt}\omega,\alpha},{\cal L}^{\hspace{1pt}\sigma}_{\hspace{1pt}\tau,\beta}\}\notag\\
 &\hspace{-3pt}=(\delta_{\omega+\alpha,\tau}-(-1)^{\sigma}\delta_{\omega+\alpha,\tau+\beta}){\cal L}^{\hspace{1pt}1+\epsilon+\sigma}_{\hspace{1pt}\omega,\alpha+\beta}
 -(-1)^{\epsilon}(\delta_{\omega,\tau}-(-1)^{\sigma}\delta_{\omega,\tau+\beta}){\cal L}^{\hspace{1pt}1+\epsilon+\sigma}_{\hspace{1pt}\omega+\alpha,\alpha+\beta}.
 \end{align}
 The commutator of Eq.~\ref{eqcommL} has compressed the commutation relations
 verbosely listed in from Eqs.~\ref{A1} to~\ref{A6}.

\appendix{~~Abelian Subalgebra Extension\label{appabelext}}
 As aforedescribed, a {\em path} on the group manifold of $SU(N)$ is depending on
 designations of subalgebras in a decomposition sequence.
 For the expositions of quotient algebras of higher ranks as well~\cite{SuTsai2,SuTsai3},
 it becomes essential to search the Cartan subalgebras of $su(N)$.
 Toward this purpose, an easy method taking advantage of quotient algebras of rank zero
 is here introduced.
 This method considers only Cartan subalgebras composed of ``{\em basis generators}" and
 ignores those in superpositions of noncommuting generators.
 The intrinsic center subalgebra ${\cal C}$ is
 always a convenient choice to start with.
 Inside an example for $su(4)$ shown in Fig.~\ref{csu4},
 the center subalgebra
 ${\cal C}=\{\sigma_3\otimes I,I\otimes\sigma_3,\sigma_3\otimes\sigma_3\}$ is placed
 in the central column of the quotient algebra. In this quotient algebra,
 each abelian subspace of the conjugate pairs can find one corresponding
 generator in ${\cal C}$ such that the subspace recovers to
 be a Cartan subalgebra by absorbing this generator.
 For instance, the Cartan subalgebra
 $\{I\otimes\sigma_1,\sigma_3\otimes\sigma_1,\sigma_3\otimes I\}$ is gained by
 taking the union of $W_1$ and $\{\sigma_3\otimes I\}\subset{\cal C}$.
 Consequently, $6$ Cartan subalgebras
 of {\em nearest neighbors} are {\em extended} from ${\cal C}$.
 They form the set of Cartan subalgebras of
 {\em the extension of the $1$st shell}.

 Similarly, the Cartan subalgebras
 of the $2$nd-shell extension are constructed from the $6$ quotient algebras given
 by the Cartan subalgebras obtained in the $1$st shell.
 Among other redundancies, there appear $8$ new subalgebras 
 in this shell and no more new member in the following shells.
 In other words, the Lie algebra $su(4)$ has $15$ Cartan subalgebras in total.
 Based on structures of quotient algebras of rank zero,
 it will be proved in~\cite{SuTsai1} that
 all the Cartan subalgebras of the Lie algebra $su(N)$, $2^{p-1}<N\leq 2^p$,
 can be found in the first $p$ shells of the extension.
 The success of the method is attributed to the key clue that, 
 within the quotient algebra $\{{\cal Q(A)}\}$ given by a Cartan subalgebra ${\cal A}$,
 there exists a commuting subset $\mathfrak{B}\subset{\cal A}$,
 {\em i.e.}, $[W,\mathfrak{B}]=0$,
 for every abelian subspace 
 $W\in\{{\cal Q(A)}\}$.
 Then, another Cartan subalgebra $W\cup\mathfrak{B}$ is extended from ${\cal A}$
 by taking the union of the subspace and the the subset $\mathfrak{B}$, which is known as
 {\em a maximal bi-subalgebra of} ${\cal A}$ in the language of the $s$-representation.

 For a Lie algebra of dimension a power of $2$,
 the identical set of Cartan subalgebras is acquired irrespective of
 with which subalgebra the extension begins.
 A reminder is that the set of Cartan subalgebras of the Lie algebra
 $su(N)$ with $2^{p-1}<N<2^p$  is obtained by applying the removing process
 to the original set of $su(2^p)$.
 Since each of its Cartan subalgebras has the same number $2(2^p-1)$ 
 of {\em nearest neighbors} and any one of these subalgebras is reachable within
 $p$ shells of the extension regardless of the starting member,
 it may be of help to envision 
 the set of Cartan subalgebras of $su(2^p)$ in terms of {\em a hypercube-like topology}.



\newpage
\appendix{~~Tables of Quotient Algebras\label{apptableQA}}

\begin{figure}[htbp]
\begin{center}
\[\begin{array}{ccc}
 \hspace{-100pt}(a)\\
 \\
 \begin{array}{c}
 I\otimes I\otimes \sigma_{1}\\
 \\
 \\
 \\
 \\
 \\
 \\
 \end{array}
 &
 \begin{array}{c}
 \hspace{-5pt}\sigma_{3}\otimes\hspace{0pt} I\hspace{5pt}\otimes I\\
 I\hspace{0pt} \otimes \sigma_{3}\otimes\hspace{0pt} I\\
 \hspace{5pt}I\hspace{0pt}\otimes\hspace{5pt} I\hspace{0pt}\otimes \sigma_{3}\\
 \hspace{-3pt}\sigma_{3}\otimes \sigma_{3}\otimes\hspace{2pt} I\\
 \hspace{0pt}\sigma_{3}\otimes\hspace{3pt} I\hspace{2pt}\otimes\hspace{0pt} \sigma_{3}\\
 \hspace{2pt}I\hspace{3pt}\otimes \sigma_{3}\otimes \sigma_{3}\\
 \sigma_{3}\otimes \sigma_{3}\otimes \sigma_{3}
 \end{array}
 &
 \begin{array}{c}
 \hspace{3pt}I\hspace{2pt}\otimes I\hspace{3pt}\otimes \sigma_{2} \\
 \hspace{1pt}\sigma_{3} \otimes\hspace{3pt} I\hspace{1pt}\otimes\hspace{0pt} \sigma_{2}  \\
 \hspace{3pt}I\hspace{2pt}\otimes \sigma_{3}\otimes \sigma_{2}   \\
 \sigma_{3}\otimes \sigma_{3}\otimes \sigma_{2}
 \\
 \\
 \\
 \\
 \end{array}
 \end{array}\]
\\
\[\begin{array}{ccc}
 \hspace{-98pt}(b)\\
\\
 \begin{array}{c}
 \hspace{3pt}I\hspace{2pt}\otimes I\hspace{3pt}\otimes \sigma_{1} \\
 \hspace{0pt}\sigma_{3} \otimes\hspace{3pt} I\hspace{1pt}\otimes\hspace{0pt} \sigma_{1}  \\
 \hspace{3pt}I\hspace{2pt}\otimes \sigma_{3}\otimes \sigma_{1}   \\
 \sigma_{3}\otimes \sigma_{3}\otimes \sigma_{1}
 \\
 \\
 \\
 \\
 \end{array}
 &
 \begin{array}{c}
 \hspace{-3pt}\sigma_{3}\otimes\hspace{2pt} I\hspace{2pt}\otimes I\\
 I\hspace{2pt} \otimes \sigma_{3}\otimes\hspace{1pt} I\\
 \hspace{3pt}I\hspace{2pt}\otimes\hspace{1pt} I\hspace{3pt}\otimes \sigma_{3}\\
 \hspace{-1pt}\sigma_{3}\otimes \sigma_{3}\otimes\hspace{3pt} I\\
 \hspace{1pt}\sigma_{3}\otimes\hspace{2pt} I\hspace{2pt}\otimes\hspace{1pt} \sigma_{3}\\
 \hspace{2pt}I\hspace{3pt}\otimes \sigma_{3}\otimes \sigma_{3}\\
 \sigma_{3}\otimes \sigma_{3}\otimes \sigma_{3}
 \end{array}
 &
 \begin{array}{c}
 \hspace{3pt}I\hspace{2pt}\otimes I\hspace{3pt}\otimes \sigma_{2} \\
 \hspace{1pt}\sigma_{3} \otimes\hspace{3pt} I\hspace{1pt}\otimes\hspace{0pt} \sigma_{2}  \\
 \hspace{3pt}I\hspace{2pt}\otimes \sigma_{3}\otimes \sigma_{2}   \\
 \sigma_{3}\otimes \sigma_{3}\otimes \sigma_{2}
 \\
 \\
 \\
 \\
 \end{array}
\end{array}\]
\end{center}
\fcaption{(a) Given the intrinsic center subalgebra ${\cal C}\subset su(8)$ listed in the
 central column, an abelian subspace $\hat{W}_1$ at the RHS produced from
 the commutator $[g,{\cal C}]$ with a seed
 $g=I\otimes I\otimes \sigma_{1}$ at the LHS;
 (b) in the reverse step, the conjugate subspace $W_1$ at the
 LHS similarly produced
 with a seed arbitrarily taken from $\hat{W}_1$. \label{stepsu8}}
\end{figure}

 \newpage
\begin{figure}
\begin{center}
\[\begin{array}{c}
 \begin{array}{c}
 \hspace{-3pt}\sigma_{3}\otimes\hspace{2pt} I\hspace{2pt}\otimes I\\
 \hspace{2pt}I\hspace{1pt} \otimes \sigma_{3}\otimes\hspace{2pt} I\\
 \hspace{4pt}I\hspace{2pt}\otimes\hspace{1pt} I\hspace{3.5pt}\otimes \sigma_{3}\\
 \hspace{-1pt}\sigma_{3}\otimes \sigma_{3}\otimes\hspace{3.5pt} I\\
 \hspace{0pt}\sigma_{3}\otimes\hspace{2pt} I\hspace{3pt}\otimes\hspace{0pt} \sigma_{3}\\
 \hspace{2pt}I\hspace{3pt}\otimes \sigma_{3}\otimes \sigma_{3}\\
 \sigma_{3}\otimes \sigma_{3}\otimes \sigma_{3}
 \end{array}
 \\
 \\
 \begin{array}{ccccccc}
  W_{1}
  &
  \begin{array}{cc}
  \hspace{1pt}I\otimes\hspace{2pt} I\hspace{2pt}\otimes \hspace{0pt}\sigma_{1}&\sigma_{3}\otimes\hspace{2pt} I\hspace{2pt}\otimes\sigma_{1}\\
  I\otimes \sigma_{3}\otimes \sigma_{1}&\sigma_{3}\otimes \sigma_{3}\otimes \sigma_{1}
  \end{array}
  &
  &
  \begin{array}{cc}
  \hspace{1pt}I\hspace{-1pt}\otimes\hspace{2pt} I\hspace{2pt}\otimes \sigma_{2}&\sigma_{3}\otimes\hspace{2pt} I\hspace{2pt}\otimes \sigma_{2}\\
  I\otimes \sigma_{3}\otimes \sigma_{2} &   \sigma_{3}\otimes \sigma_{3}\otimes \sigma_{2}
  \end{array}
  &
  \hat{W}_{1}
  \\
  \\
  W_{2}
  &
  \begin{array}{cc}
  \hspace{-2pt}I\otimes \sigma_{1}\otimes \hspace{2pt}I  &   \hspace{-2pt}\sigma_{3}\otimes \sigma_{1}\otimes\hspace{2pt} I \\
  I\otimes \sigma_{1}\otimes \sigma_{3}  &  \sigma_{3}\otimes \sigma_{1}\otimes \sigma_{3}
  \end{array}
  &
  &
  \begin{array}{cc}
  \hspace{-3pt}I\otimes \sigma_{2}\otimes\hspace{1pt} I  &  \hspace{-3pt}\sigma_{3}\otimes \sigma_{2}\otimes\hspace{1pt} I  \\
  I\otimes \sigma_{2}\otimes \sigma_{3}  &  \sigma_{3}\otimes \sigma_{2}\otimes \sigma_{3}
  \end{array}
  &
  \hat{W}_{2}
  \\
  \\
  W_{3}
  &
  \begin{array}{cc}
  \hspace{3pt}I\hspace{2pt}\otimes \sigma_{1}\otimes \sigma_{1}  &  \hspace{3pt}I\hspace{2pt}\otimes \sigma_{2}\otimes \sigma_{2}  \\
  \sigma_{3}\otimes \sigma_{1}\otimes \sigma_{1}  &  \sigma_{3}\otimes \sigma_{2}\otimes \sigma_{2}
  \end{array}
  &
  &
  \begin{array}{cc}
  \hspace{3pt}I\hspace{2pt}\otimes \sigma_{2}\otimes \sigma_{1}  &  \hspace{3pt}I\hspace{2pt}\otimes \sigma_{1}\otimes \sigma_{2} \\
   \sigma_{3}\otimes \sigma_{2}\otimes \sigma_{1}  &  \sigma_{3}\otimes \sigma_{1}\otimes\sigma_{2}
  \end{array}
  &
  \hat{W}_{3}
  \\
  \\
  W_{4}
  &
  \begin{array}{cc}
  \hspace{-2pt}\sigma_{1}\otimes\hspace{1pt} I\otimes\hspace{2pt} I  &  \hspace{-2pt}\sigma_{1}\otimes \sigma_{3}\otimes\hspace{2pt} I   \\
  \sigma_{1}\otimes I\otimes \sigma_{3}  &  \sigma_{1}\otimes \sigma_{3}\otimes \sigma_{3}
  \end{array}
  &
  &
  \begin{array}{cc}
  \hspace{-2pt}\sigma_{2}\otimes\hspace{0pt} I\otimes\hspace{2pt} I  &   \hspace{-2pt}\sigma_{2}\otimes \sigma_{3}\otimes\hspace{2pt} I  \\
  \sigma_{2}\otimes I\otimes \sigma_{3}  &   \sigma_{2}\otimes \sigma_{3}\otimes \sigma_{3}
  \end{array}
  &
  \hat{W}_{4}
  \\
  \\
  W_{5}
  &
  \begin{array}{cc}
  \sigma_{1}\otimes\hspace{2pt} I\hspace{3pt}\otimes \sigma_{1}  &  \sigma_{2}\otimes \hspace{2pt}I\hspace{3pt}\otimes \sigma_{2}  \\
  \sigma_{1}\otimes \sigma_{3}\otimes \sigma_{1}  &  \sigma_{2}\otimes \sigma_{3}\otimes \sigma_{2}
  \end{array}
  &
  &
  \begin{array}{cc}
  \sigma_{2}\otimes\hspace{2pt} I\hspace{3pt}\otimes \sigma_{1}  &  \sigma_{1}\otimes\hspace{2pt} I\hspace{3pt}\otimes \sigma_{2} \\
  \sigma_{2}\otimes \sigma_{3}\otimes \sigma_{1}  &  \sigma_{1}\otimes \sigma_{3}\otimes \sigma_{2}
  \end{array}
  &
  \hat{W}_{5}
  \\
  \\
  W_{6}
  &
  \begin{array}{cc}
  \hspace{-3pt}\sigma_{1}\otimes \sigma_{1}\otimes \hspace{2pt}I  &  \hspace{-2pt}\sigma_{2}\otimes \sigma_{2}\otimes\hspace{2pt} I   \\
  \sigma_{1}\otimes \sigma_{1}\otimes \sigma_{3}  &  \sigma_{2}\otimes \sigma_{2}\otimes \sigma_{3}
  \end{array}
  &
  &
  \begin{array}{cc}
  \hspace{-2pt}\sigma_{2}\otimes \sigma_{1}\otimes\hspace{2pt} I  &  \hspace{-3pt}\sigma_{1}\otimes \sigma_{2}\otimes\hspace{2pt} I \\
  \sigma_{2}\otimes \sigma_{1}\otimes \sigma_{3}  &  \sigma_{1}\otimes \sigma_{2}\otimes \sigma_{3}
  \end{array}
  &
  \hat{W}_{6}
  \\
  \\
  W_{7}
  &
  \begin{array}{cc}
  \sigma_{1}\otimes \sigma_{1}\otimes \sigma_{1}  &  \sigma_{2}\otimes \sigma_{2}\otimes \sigma_{1}   \\
  \sigma_{2}\otimes \sigma_{1}\otimes \sigma_{2}  &  \sigma_{1}\otimes \sigma_{2}\otimes \sigma_{2}
  \end{array}
  &
  &
  \begin{array}{cc}
   \sigma_{2}\otimes \sigma_{1}\otimes \sigma_{1}  &  \sigma_{1}\otimes \sigma_{2}\otimes \sigma_{1} \\
   \sigma_{1}\otimes \sigma_{1}\otimes \sigma_{2}  &  \sigma_{2}\otimes \sigma_{2}\otimes \sigma_{2}
  \end{array}
  &
  \hat{W}_{7}
  \\
  \end{array}
  \end{array} \]
  \end{center}
  \fcaption{The quotient algebra of rank zero given by the intrinsic Cartan subalgebra of $su(8)$.\label{csu8}}
\end{figure}

\newpage
\begin{figure}
\begin{center}
\[\begin{array}{c}
  \begin{array}{c}
  \hspace{2pt}I\hspace{3pt}\otimes \sigma_{3}\\
  \hspace{-2pt}\mu_{3} \otimes \hspace{3pt}I  \\
  \hspace{-2pt}\mu_{8} \otimes\hspace{3pt} I  \\
  \mu_{3} \otimes \sigma_{3} \\
  \mu_{8} \otimes \sigma_{3}
  \end{array}
  \\
  \\
 \begin{array}{ccccc}

 W_{1}
 &
 \begin{array}{c}
 I \otimes \sigma_{1}\hspace{5pt}\mu_{3}\otimes \sigma_{1}\hspace{5pt}\mu_{8}\otimes \sigma_{1}
 \end{array}
 &
 &
 \begin{array}{c}
 I\otimes \sigma_{2}\hspace{5pt}\mu_{3}\otimes \sigma_{2}\hspace{5pt}\mu_{8} \otimes \sigma_{2}
 \end{array}
 &
 \hat{W}_{1}
 \\
 \\
 W_{2}
 &
 \begin{array}{c}
 \mu_{1} \otimes\hspace{2pt} I \hspace{13pt} \mu_{1}\otimes\sigma_{3}
 \end{array}
 &
 &
 \begin{array}{c}
 \mu_{2} \otimes\hspace{2pt} I  \hspace{13pt}\mu_{2} \otimes \sigma_{3}
 \end{array}
 &
 \hat{W}_{2}
  \\
  \\
 W_{3}
 &
 \begin{array}{c}
 \mu_{1} \otimes \sigma_{1} \hspace{10pt} \mu_{2}\otimes \sigma_{2}
 \end{array}
 &
 &
 \begin{array}{c}
 \mu_{1}  \otimes \sigma_{2} \hspace{10pt} \mu_{2}\otimes \sigma_{1}
 \end{array}
 &
 \hat{W}_{3}
 \\
 \\
 W_{4}
 &
 \begin{array}{c}
 \mu_{4} \otimes \hspace{2pt}I\hspace{13pt}\mu_{4}\otimes \sigma_{3}
 \end{array}
 &
 &
 \begin{array}{c}
 \mu_{5} \otimes\hspace{2pt} I\hspace{13pt}\mu_{5} \otimes \sigma_{3}
 \end{array}
 &
 \hat{W}_{4}
 \\
 \\
 W_{5}
 &
 \begin{array}{c}
 \mu_{4} \otimes \sigma_{1}\hspace{10pt} \mu_{5}\otimes \sigma_{2}
 \end{array}
 &
 &
 \begin{array}{c}
 \mu_{4} \otimes \sigma_{2}\hspace{10pt} \mu_{5}\otimes \sigma_{1}
 \end{array}
 &
 \hat{W}_{5}
 \\
  \\
 W_{6}
 &
  \begin{array}{c}
 \mu_{6} \otimes\hspace{2pt} I \hspace{13pt} \mu_{6}\otimes \sigma_{3}
 \end{array}
 &
 &
  \begin{array}{c}
  \mu_{7} \otimes\hspace{2pt} I \hspace{13pt}\mu_{7}\otimes \sigma_{3}
 \end{array}
 &
 \hat{W}_{6}
 \\
   \\
 W_{7}
 &
 \begin{array}{c}
  \mu_{6} \otimes  \sigma_{1}\hspace{10pt} \mu_{7}\otimes \sigma_{2}
 \end{array}
 &
 &
 \begin{array}{c}
  \mu_{6}\otimes  \sigma_{2}\hspace{10pt} \mu_{7}\otimes \sigma_{1}
 \end{array}
 &
 \hat{W}_{7}
 \end{array}
 \end{array}\]
 \fcaption{The quotient algebra of rank zero given by the intrinsic Cartan subalgebra of $su(6)$, where
  the generators $\mu_j$ denoting the Gell-Mann matrices and the former $I$, before the symbol $\otimes$,
  being the $3\times 3$ identity in contrast to the $2\times 2$ identity of the latter.\label{csu6}}
\end{center}
\end{figure}

\newpage
\begin{figure}
\begin{center}
\[\begin{array}{c}
  \begin{array}{c}
\hspace{-4pt}\sigma_{1}\otimes\hspace{5pt} I\otimes I   \\
\hspace{2pt}I\hspace{1pt} \otimes \sigma_{1}\otimes\hspace{3pt} I    \\
\hspace{4pt}I\hspace{1pt}\otimes I\hspace{4pt}\otimes \sigma_{1}   \\
\hspace{-3pt}\sigma_{1}\otimes \sigma_{1}\otimes\hspace{3pt} I  \\
\hspace{-1pt}\sigma_{1}\otimes\hspace{3pt} I\hspace{2pt}\otimes \sigma_{1}  \\
\hspace{3pt}I\hspace{1pt}\otimes \sigma_{1}\otimes \sigma_{1}  \\
\sigma_{1}\otimes \sigma_{1}\otimes \sigma_{1}
  \end{array}
  \\
  \\
\begin{array}{ccccc}
 W_{1}
 &
 \begin{array}{cc}
 \hspace{-2pt}\sigma_{3}\otimes I\otimes\hspace{3pt} I &\hspace{-2pt}\sigma_{3}\otimes \sigma_{1}\otimes\hspace{3pt} I\\
 \sigma_{3}\otimes I\otimes \sigma_{1}  &  \sigma_{3}\otimes \sigma_{1}\otimes \sigma_{1}
 \end{array}
 &
 &
 \begin{array}{cc}
 \hspace{-2pt}\sigma_{2}\otimes I\otimes \hspace{3pt}I  &   \hspace{-2pt}\sigma_{2}\otimes \sigma_{1}\otimes\hspace{3pt} I  \\
 \sigma_{2}\otimes I\otimes \sigma_{1}  &   \sigma_{2}\otimes \sigma_{1}\otimes \sigma_{1}
 \end{array}
 &
 \hat{W}_{1}
 \\
 \\
 W_{2}
 &
 \begin{array}{cc}
 \hspace{-2pt}I\otimes \sigma_{3}\otimes\hspace{3pt} I  &  \hspace{-2pt}\sigma_{1}\otimes \sigma_{3}\otimes\hspace{3pt} I  \\
 I\otimes \sigma_{3}\otimes \sigma_{1}  &  \sigma_{1}\otimes \sigma_{3}\otimes \sigma_{1}
 \end{array}
 &
 &
 \begin{array}{cc}
 \hspace{-2pt}I\otimes \sigma_{2}\otimes\hspace{3pt} I  &  \hspace{-2pt}\sigma_{1}\otimes \sigma_{2}\otimes\hspace{3pt} I  \\
 I\otimes \sigma_{2}\otimes \sigma_{1}  &   \sigma_{1}\otimes \sigma_{2}\otimes \sigma_{1}
 \end{array}
 &
 \hat{W}_{2}
 \\
 \\
 W_{3}
 &
 \begin{array}{cc}
 \hspace{-1pt}I\hspace{1pt}\otimes\hspace{1pt} I\hspace{4pt}\otimes \sigma_{3} &  \sigma_{1}\otimes\hspace{3pt} I\hspace{2pt}\otimes \sigma_{3}\\
 I\otimes \sigma_{1}\otimes \sigma_{3} &   \sigma_{1}\otimes \sigma_{1}\otimes \sigma_{3}
 \end{array}
 &
 &
 \begin{array}{cc}
 \hspace{-1pt}I\hspace{1pt}\otimes\hspace{1pt} I\hspace{4pt}\otimes \sigma_{2}  &   \sigma_{1} \otimes\hspace{3pt} I\hspace{2pt}\otimes \sigma_{2}\\
 I\otimes \sigma_{1}\otimes \sigma_{2} &   \sigma_{1}\otimes \sigma_{1}\otimes \sigma_{2}
 \end{array}
 &
 \hat{W}_{3}
 \\
 \\
 W_{4}
 &
 \begin{array}{cc}
 \hspace{-2pt}\sigma_{3}\otimes \sigma_{3}\otimes\hspace{3pt} I  &  \hspace{-2pt}\sigma_{2}\otimes \sigma_{2}\otimes \hspace{3pt}I\\
 \sigma_{3}\otimes \sigma_{3}\otimes \sigma_{1}  &   \sigma_{2}\otimes \sigma_{2}\otimes \sigma_{1}
 \end{array}
 &
 &
 \begin{array}{cc}
 \hspace{-2pt}\sigma_{2}\otimes \sigma_{3}\otimes\hspace{3pt} I   & \hspace{-2pt} \sigma_{3}\otimes \sigma_{2}\otimes\hspace{3pt} I \\
 \sigma_{2}\otimes \sigma_{3}\otimes \sigma_{1}   & \sigma_{3}\otimes \sigma_{2}\otimes \sigma_{1}
 \end{array}
 &
 \hat{W}_{4}
 \\
 \\
 W_{5}
 &
 \begin{array}{cc}
 \sigma_{3}\otimes\hspace{2pt} I\hspace{3pt}\otimes \sigma_{3}  &   \sigma_{2}\otimes\hspace{2pt} I\hspace{3pt}\otimes \sigma_{2} \\
 \sigma_{3}\otimes \sigma_{1}\otimes \sigma_{3}  &   \sigma_{2}\otimes \sigma_{1}\otimes \sigma_{2}
 \end{array}
 &
 &
 \begin{array}{cc}
 \sigma_{2}\otimes\hspace{2pt} I\hspace{3pt}\otimes \sigma_{3}  &  \sigma_{3}\otimes\hspace{2pt} I\hspace{3pt}\otimes \sigma_{2} \\
 \sigma_{2}\otimes \sigma_{1}\otimes \sigma_{3}  &   \sigma_{3}\otimes \sigma_{1}\otimes \sigma_{2}
 \end{array}
 &
 \hat{W}_{5}
 \\
 \\
 W_{6}
 &
 \begin{array}{cc}
 \hspace{5pt}I\otimes \sigma_{3}\otimes \sigma_{3}  &   \hspace{5pt}I\otimes \sigma_{2}\otimes \sigma_{2}\\
 \sigma_{1}\otimes \sigma_{3}\otimes \sigma_{3}  &  \sigma_{1}\otimes \sigma_{2}\otimes \sigma_{2}
 \end{array}
 &
 &
 \begin{array}{cc}
 \hspace{5pt}I\otimes \sigma_{2}\otimes \sigma_{3}   &  \hspace{5pt}I\otimes \sigma_{3}\otimes \sigma_{2} \\
 \sigma_{1}\otimes \sigma_{2}\otimes \sigma_{3}   &  \sigma_{1}\otimes \sigma_{3}\otimes \sigma_{2}
 \end{array}
 &
 \hat{W}_{6}
 \\
 \\
 W_{7}
 &
 \begin{array}{cc}
 \sigma_{3}\otimes \sigma_{3}\otimes \sigma_{3}   &  \sigma_{2}\otimes \sigma_{2}\otimes \sigma_{3}\\
 \sigma_{2}\otimes \sigma_{3}\otimes \sigma_{2}  & \sigma_{3}\otimes \sigma_{2}\otimes \sigma_{2}
 \end{array}
 &
 &
 \begin{array}{cc}
 \sigma_{2}\otimes \sigma_{3}\otimes \sigma_{3}   &  \sigma_{3}\otimes \sigma_{2}\otimes \sigma_{3} \\

 \sigma_{3}\otimes \sigma_{3}\otimes \sigma_{2}  &  \sigma_{2}\otimes \sigma_{2}\otimes \sigma_{2}
 \end{array}
 &
 \hat{W}_{7}
 \end{array}
 \end{array}\]
 \end{center}
 \fcaption{The quotient algebra of rank zero given by a non-diagonal Cartan subalgebra of $su(8)$.\label{ncsu8}}
\end{figure}

\newpage
\begin{figure}
\begin{center}
\[\begin{array}{c}
  \begin{array}{c}
  \hspace{2pt}I\hspace{2pt}\otimes \sigma_{1}  \\
  \hspace{-3pt}\mu_{1} \otimes\hspace{2pt} I    \\
  \hspace{-3pt}\mu_{8} \otimes\hspace{2pt} I  \\
  \mu_{1} \otimes \sigma_{1} \\
  \mu_{8} \otimes \sigma_{1}
 \end{array}
 \\
 \\
\begin{array}{ccccc}
 W_{1}
 &
 \begin{array}{c}
  I \otimes \sigma_{3}\hspace{5pt}\mu_{1}\otimes \sigma_{3}\hspace{5pt}\mu_{8}\otimes \sigma_{3}
 \end{array}
 &
 &
 \begin{array}{c}
 I\otimes \sigma_{2}\hspace{5pt}\mu_{1}\otimes \sigma_{2}\hspace{5pt}\mu_{8} \otimes \sigma_{2}
 \end{array}
 &
 \hat{W}_{1}
 \\
 \\
 W_{2}
 &
 \begin{array}{c}
  \mu_{3} \otimes I\hspace{15pt}\mu_{3}\otimes\sigma_{1}
 \end{array}
 &
 &
 \begin{array}{c}
 \mu_{2} \otimes I\hspace{15pt}\mu_{2} \otimes \sigma_{1}
 \end{array}
 &
 \hat{W}_{2}
 \\
 \\
 W_{3}
 &
 \begin{array}{c}
 \mu_{4} \otimes I\hspace{15pt}\mu_{4}\otimes \sigma_{1}
 \end{array}
 &
 &
 \begin{array}{c}
  \mu_{7} \otimes I\hspace{15pt}\mu_{7} \otimes \sigma_{1}
 \end{array}
 &
 \hat{W}_{3}
 \\
 \\
 W_{4}
 &
 \begin{array}{c}
  \hspace{0pt}\mu_{6} \otimes \hspace{0pt}I\hspace{15pt}\mu_{6}\otimes \sigma_{1}
 \end{array}
 &
 &
 \begin{array}{c}
  \mu_{5} \otimes I\hspace{15pt} \mu_{5}\otimes \sigma_{1}
 \end{array}
 &
 \hat{W}_{4}
 \\
 \\
 W_{5}
 &
 \begin{array}{c}
  \mu_{3} \otimes \sigma_{3}\hspace{10pt} \mu_{2}\otimes \sigma_{2}
 \end{array}
 &
 &
 \begin{array}{c}
  \mu_{3}  \otimes \sigma_{2}\hspace{10pt}\mu_{2}\otimes \sigma_{3}
 \end{array}
 &
 \hat{W}_{5}
 \\
   & &  \\
 W_{6}
 &
 \begin{array}{c}
  \mu_{4} \otimes \sigma_{3}\hspace{10pt}\mu_{7}\otimes \sigma_{2}
 \end{array}
 &
 &
 \begin{array}{c}
  \mu_{4} \otimes \sigma_{2}\hspace{10pt}\mu_{7}\otimes \sigma_{3}
 \end{array}
 &
 \hat{W}_{6}
 \\
 \\
 W_{7}
 &
 \begin{array}{c}
  \mu_{6} \otimes  \sigma_{3}\hspace{10pt}\mu_{5}\otimes \sigma_{2}
 \end{array}
 &
 &
 \begin{array}{c}
   \mu_{6}\otimes  \sigma_{2}\hspace{10pt}\mu_{5}\otimes \sigma_{3}
 \end{array}
 &
 \hat{W}_{7}
 \end{array}
 \end{array}\]
 \fcaption{The quotient algebra of rank zero given by a non-diagonal Cartan subalgebra of $su(6)$.\label{ncsu6}}
\end{center}
\end{figure}

\newpage
\begin{figure}
\begin{center}
\[\begin{array}{c}
\begin{array}{c}
 diag\{1, 1, 1, 1,-1,-1,-1,-1 \}  \\
 diag\{1, 1,-1,-1, 1, 1,-1,-1 \} \\
 diag\{1,-1, 1,-1, 1,-1, 1,-1 \} \\
 diag\{1, 1,-1,-1,-1,-1, 1, 1 \} \\
 diag\{1,-1, 1,-1,-1, 1,-1, 1 \} \\
 diag\{1,-1,-1, 1, 1,-1,-1, 1 \} \\
 diag\{1,-1,-1, 1,-1, 1, 1,-1 \}
 \end{array}\\
 \\
 \begin{array}{ccccc}
 W_{001}
 &\begin{array}{c}
   \lambda_{12}+\lambda_{34}+\lambda_{56}+\lambda_{78}\\
   \lambda_{12}+\lambda_{34}-\lambda_{56}-\lambda_{78}\\
   \lambda_{12}-\lambda_{34}+\lambda_{56}-\lambda_{78}\\
   \lambda_{12}-\lambda_{34}-\lambda_{56}+\lambda_{78}
  \end{array}
 &
 \hspace{30pt}
 &
  \begin{array}{c}
   \hat{\lambda}_{12}+\hat{\lambda}_{34}+\hat{\lambda}_{56}+\hat{\lambda}_{78}\\
   \hat{\lambda}_{12}+\hat{\lambda}_{34}-\hat{\lambda}_{56}-\hat{\lambda}_{78}\\
   \hat{\lambda}_{12}-\hat{\lambda}_{34}+\hat{\lambda}_{56}-\hat{\lambda}_{78}\\
   \hat{\lambda}_{12}-\hat{\lambda}_{34}-\hat{\lambda}_{56}+\hat{\lambda}_{78}
  \end{array}
 & \hat{W}_{001}
 \\
 \\
 W_{010}
 &
 \begin{array}{c}
 \lambda_{13}+\lambda_{24}+\lambda_{57}+\lambda_{68}\\
 \lambda_{13}+\lambda_{24}-\lambda_{57}-\lambda_{68}\\
 \lambda_{13}-\lambda_{24}+\lambda_{57}-\lambda_{68}\\
 \lambda_{13}-\lambda_{24}-\lambda_{57}+\lambda_{68}
 \end{array}
 &
 &
 \begin{array}{c}
 \hat{\lambda}_{13}+\hat{\lambda}_{24}+\hat{\lambda}_{57}+\hat{\lambda}_{68}\\
 \hat{\lambda}_{13}+\hat{\lambda}_{24}-\hat{\lambda}_{57}-\hat{\lambda}_{68}\\
 \hat{\lambda}_{13}-\hat{\lambda}_{24}+\hat{\lambda}_{57}-\hat{\lambda}_{68}\\
 \hat{\lambda}_{13}-\hat{\lambda}_{24}-\hat{\lambda}_{57}+\hat{\lambda}_{68}
 \end{array}
 & \hat{W}_{010}
 \\
 \\
 W_{011}
 &
 \begin{array}{c}
 \lambda_{14}+\lambda_{23}+\lambda_{58}+\lambda_{67}\\
 -\lambda_{14}+\lambda_{23}-\lambda_{58}+\lambda_{67}\\
 \lambda_{14}+\lambda_{23}-\lambda_{58}-\lambda_{67}\\
 -\lambda_{14}+\lambda_{23}+\lambda_{58}-\lambda_{67}
 \end{array}
 &
 &
 \begin{array}{c}
 \hat{\lambda}_{14}+\hat{\lambda}_{23}+\hat{\lambda}_{58}+\hat{\lambda}_{67}\\
 \hat{\lambda}_{14}-\hat{\lambda}_{23}+\hat{\lambda}_{58}-\hat{\lambda}_{67}\\
 \hat{\lambda}_{14}+\hat{\lambda}_{23}-\hat{\lambda}_{58}-\hat{\lambda}_{67}\\
 \hat{\lambda}_{14}-\hat{\lambda}_{23}-\hat{\lambda}_{58}+\hat{\lambda}_{67}
 \end{array}
 & \hat{W}_{011}
 \\
 \\
 W_{100}
 &
 \begin{array}{c}
 \lambda_{15}+\lambda_{26}+\lambda_{37}+\lambda_{48}\\
 \lambda_{15}+\lambda_{26}-\lambda_{37}-\lambda_{48}\\
 \lambda_{15}-\lambda_{26}+\lambda_{37}-\lambda_{48}\\
 \lambda_{15}-\lambda_{26}-\lambda_{37}+\lambda_{48}
 \end{array}
 &
 &
 \begin{array}{c}
 \hat{\lambda}_{15}+\hat{\lambda}_{26}+\hat{\lambda}_{37}+\hat{\lambda}_{48}\\
 \hat{\lambda}_{15}+\hat{\lambda}_{26}-\hat{\lambda}_{37}-\hat{\lambda}_{48}\\
 \hat{\lambda}_{15}-\hat{\lambda}_{26}+\hat{\lambda}_{37}-\hat{\lambda}_{48}\\
 \hat{\lambda}_{15}-\hat{\lambda}_{26}-\hat{\lambda}_{37}+\hat{\lambda}_{48}
 \end{array}
 & \hat{W}_{100}
 \\
 \\
 W_{101}
 &
 \begin{array}{c}
 \lambda_{16}+\lambda_{25}+\lambda_{38}+\lambda_{47}\\
 -\lambda_{16}+\lambda_{25}-\lambda_{38}+\lambda_{47}\\
 \lambda_{16}+\lambda_{25}-\lambda_{38}-\lambda_{47}\\
 -\lambda_{16}+\lambda_{25}+\lambda_{38}-\lambda_{47}
 \end{array}
 &
 &
 \begin{array}{c}
 \hat{\lambda}_{16}+\hat{\lambda}_{25}+\hat{\lambda}_{38}+\hat{\lambda}_{47}\\
 \hat{\lambda}_{16}-\hat{\lambda}_{25}+\hat{\lambda}_{38}-\hat{\lambda}_{47}\\
 \hat{\lambda}_{16}+\hat{\lambda}_{25}-\hat{\lambda}_{38}-\hat{\lambda}_{47}\\
 \hat{\lambda}_{16}-\hat{\lambda}_{25}-\hat{\lambda}_{38}+\hat{\lambda}_{47}
 \end{array}
 & \hat{W}_{101}
 \\
 \\
 W_{110}
 &
 \begin{array}{c}
 \lambda_{17}+\lambda_{28}+\lambda_{35}+\lambda_{46}\\
 -\lambda_{17}-\lambda_{28}+\lambda_{35}+\lambda_{46}\\
 \lambda_{17}-\lambda_{28}+\lambda_{35}-\lambda_{46}\\
 -\lambda_{17}+\lambda_{28}+\lambda_{35}-\lambda_{46}
 \end{array}
 &
 &
 \begin{array}{c}
 \hat{\lambda}_{17}+\hat{\lambda}_{28}+\hat{\lambda}_{35}+\hat{\lambda}_{46}\\
 \hat{\lambda}_{17}+\hat{\lambda}_{28}-\hat{\lambda}_{35}-\hat{\lambda}_{46}\\
 \hat{\lambda}_{17}-\hat{\lambda}_{28}+\hat{\lambda}_{35}-\hat{\lambda}_{46}\\
 \hat{\lambda}_{17}-\hat{\lambda}_{28}-\hat{\lambda}_{35}+\hat{\lambda}_{46}
 \end{array}
 &
 \hat{W}_{110}
 \\
 \\
 W_{111}
 &
 \begin{array}{c}
 \lambda_{18}+\lambda_{27}+\lambda_{36}+\lambda_{45}\\
 -\lambda_{18}-\lambda_{27}+\lambda_{36}+\lambda_{45}\\
 -\lambda_{18}+\lambda_{27}-\lambda_{36}+\lambda_{45}\\
 -\lambda_{18}+\lambda_{27}+\lambda_{36}-\lambda_{45}
 \end{array}
 &
 &
 \begin{array}{c}
 \hat{\lambda}_{18}+\hat{\lambda}_{27}+\hat{\lambda}_{36}+\hat{\lambda}_{45}\\
 \hat{\lambda}_{18}+\hat{\lambda}_{27}-\hat{\lambda}_{36}-\hat{\lambda}_{45}\\
 \hat{\lambda}_{18}-\hat{\lambda}_{27}+\hat{\lambda}_{36}-\hat{\lambda}_{45}\\
 -\hat{\lambda}_{18}-\hat{\lambda}_{27}+\hat{\lambda}_{36}+\hat{\lambda}_{45}
 \end{array}
 &
 \hat{W}_{111}
 \end{array}
\end{array}\]
\fcaption{The quotient algebra of Fig.\protect\ref{csu8}
 in the $\lambda$-representation.\label{Gsu8}}
\end{center}
\end{figure}

\newpage
\begin{figure}
\begin{center}
\[\begin{array}{c}
 \begin{array}{c}
 diag\{1,-1, 1,-1,1,-1 \} \\
 \hspace{-6pt}diag\{1, 1,-1,-1, 0,0 \} \\
 \hspace{-6pt}diag\{1,1, 1,1, -2,-2 \} \\
 \hspace{-6pt}diag\{1, -1,-1,1,0,0 \}  \\
 \hspace{1pt}diag\{1,-1, 1,-1,-2,2 \}
 \end{array}
 \\
 \\

\begin{array}{ccccc}

 W_{001}
 &
 \begin{array}{c}
 \lambda_{12}-\lambda_{34}\\
 \lambda_{12}+\lambda_{34}+\lambda_{56}\\
 \lambda_{12}+\lambda_{34}-2\lambda_{56}
 \end{array}
 &
 \hspace{30pt}
 &
 \begin{array}{c}
 \hat{\lambda}_{12}-\hat{\lambda}_{34}\\
 \hat{\lambda}_{12}+\hat{\lambda}_{34}+\hat{\lambda}_{56}\\
 \hat{\lambda}_{12}+\hat{\lambda}_{34}-2\hat{\lambda}_{56}
 \end{array}
 &
 \hat{W}_{001}
 \\
 \\
 W_{010}
 &
 \begin{array}{c}
 \lambda_{13}+\lambda_{24}\\
 \lambda_{13}-\lambda_{24}
 \end{array}
 &
 &
 \begin{array}{c}
 \hat{\lambda}_{13}+\hat{\lambda}_{24}\\
 \hat{\lambda}_{13}-\hat{\lambda}_{24}
 \end{array}
 &
 \hat{W}_{010}
 \\
 \\
 W_{011}
 &
 \begin{array}{c}
 \lambda_{14}+\lambda_{23}\\
 -\lambda_{14}+\lambda_{23}
 \end{array}
 &
 &
 \begin{array}{c}
 \hat{\lambda}_{14}-\hat{\lambda}_{23}\\
 \hat{\lambda}_{14}+\hat{\lambda}_{23}
 \end{array}
 &
 \hat{W}_{011}
 \\
 \\
 W_{100}
 &
 \begin{array}{c}
 \lambda_{15}+\lambda_{26}\\
 \lambda_{15}-\lambda_{26}
 \end{array}
 &
 &
 \begin{array}{c}
 \hat{\lambda}_{15}+\hat{\lambda}_{26}\\
 \hat{\lambda}_{15}-\hat{\lambda}_{26}
 \end{array}
 &
 \hat{W}_{100}
 \\
                                                                                     \\
 W_{101}
 &
 \begin{array}{c}
 \lambda_{16}+\lambda_{25}\\
 -\lambda_{16}+\lambda_{25}
 \end{array}
 &
 &
 \begin{array}{c}
 \hat{\lambda}_{16}+\hat{\lambda}_{25} \\
 \hat{\lambda}_{16}-\hat{\lambda}_{25}
 \end{array}
 &
 \hat{W}_{101}
 \\
 \\
 W_{110}
 &
 \begin{array}{c}
 \lambda_{35}+\lambda_{46} \\
 \lambda_{35}-\lambda_{46}
 \end{array}
 &
 &
 \begin{array}{c}
 \hat{\lambda}_{35}+\hat{\lambda}_{46}\\
 \hat{\lambda}_{35}-\hat{\lambda}_{46}
 \end{array}
 &
 \hat{W}_{110}
 \\
 \\
 W_{111}
 &
 \begin{array}{c}
 \lambda_{36}+\lambda_{45}\\
 -\lambda_{36}+\lambda_{45}
 \end{array}
 &
 &
 \begin{array}{c}
 \hat{\lambda}_{36}+\hat{\lambda}_{45} \\
 \hat{\lambda}_{36}-\hat{\lambda}_{45}
 \end{array}
 &
 \hat{W}_{111}
 \end{array}
 \end{array}\]
 \fcaption{The quotient algebra of Fig.\protect\ref{csu6} in the $\lambda$-representation.\label{Gsu6}}
 \end{center}
 \end{figure}

\newpage
\begin{figure}
\begin{center}
\[\begin{array}{cccccccccc}
& & \triangle_{12} & & \triangle_{34} & & \triangle_{56} & & \triangle_{78}&\\
W_{001} \text{ or } \hat{W}_{001}& & 1 & 2 & 3 & 4 & 5 & 6 & 7 & 8\\
\begin{picture}(100,30)(34,0)
\qbezier(163,30)(175,20)(185,30)\qbezier(205,30)(216,20)(227,30)\qbezier(247,30)(258,20)(269,30)\qbezier(288,30)(299,20)(310,30)
\end{picture}
& & \triangle_{13} & & \triangle_{24} & & \triangle_{57} & & \triangle_{68}&\\
W_{010} \text{ or } \hat{W}_{010} & & 1 & 2 & 3 & 4 & 5 & 6 & 7 & 8\\
\begin{picture}(100,30)(34,0)
\qbezier(163,30)(184,20)(205,30)\qbezier(185,30)(206,20)(227,30)\qbezier(247,30)(268,20)(288,30)\qbezier(269,30)(290,20)(310,30)
\end{picture}
& & \triangle_{14} & & \triangle_{23} & & \triangle_{58} & & \triangle_{67}&\\
W_{011} \text{ or } \hat{W}_{011} & & 1 & 2 & 3 & 4 & 5 & 6 & 7 & 8\\
\begin{picture}(100,30)(34,0)
\qbezier(163,30)(195,12)(227,30)\qbezier(185,30)(195,20)(205,30)\qbezier(247,30)(279,12)(310,30)\qbezier(269,30)(279,20)(288,30)
\end{picture}
& & \triangle_{15} & & \triangle_{26} & & \triangle_{38} & & \triangle_{47}&\\
W_{100} \text{ or } \hat{W}_{100} & & 1 & 2 & 3 & 4 & 5 & 6 & 7 & 8\\
\begin{picture}(100,30)(34,0)
\qbezier(163,30)(205,15)(247,30)\qbezier(185,30)(227,16)(269,30)\qbezier(205,30)(247,17)(288,30)\qbezier(227,30)(270,16)(310,30)
\end{picture}
& & \triangle_{16} & & \triangle_{25} & & \triangle_{37} & & \triangle_{48}&\\
W_{101} \text{ or } \hat{W}_{101} & & 1 & 2 & 3 & 4 & 5 & 6 & 7 & 8\\
\begin{picture}(100,30)(34,0)
\qbezier(163,30)(216,10)(269,30)\qbezier(185,30)(216,20)(247,30)\qbezier(205,30)(257,10)(310,30)\qbezier(227,30)(259,20)(288,30)
\end{picture}
& & \triangle_{17} & & \triangle_{28} & & \triangle_{35} & & \triangle_{46}&\\
W_{110} \text{ or } \hat{W}_{110} & & 1 & 2 & 3 & 4 & 5 & 6 & 7 & 8 \\
\begin{picture}(100,30)(34,0)
\qbezier(163,30)(225,10)(288,30)\qbezier(185,30)(247,10)(310,30)\qbezier(205,30)(226,20)(247,30)\qbezier(227,30)(248,20)(269,30)
\end{picture}
& & \triangle_{18} & & \triangle_{27} & & \triangle_{36} & & \triangle_{45}&\\
W_{111} \text{ or } \hat{W}_{111} & & 1 & 2 & 3 & 4 & 5 & 6 & 7 &
8
\\
\begin{picture}(100,30)(34,0)
\qbezier(163,30)(236,2)(310,30)\qbezier(185,30)(235,10)(288,30)\qbezier(205,30)(237,16)(269,30)\qbezier(227,30)(237,22)(247,30)
\end{picture}
\end{array}\]\\
\fcaption{Dividing the off-diagonal $\lambda$-generators of $su(8)$ into $7$
 conjugate pairs
 according to the binary partitioning, where
 the symbol $\triangle$ representing either a generator $\lambda_{ij}$ or a $\hat{\lambda}_{ij}$;
 there holding the condition of closure for these abelian subspaces,
 $\forall\hspace{2pt}\zeta,\eta\in Z^3_2-\{\bf 0\}$,
 $[W_{\zeta},W_{\eta}]\subset \hat{W}_{\zeta+\eta}$,
 $[W_{\zeta},\hat{W}_{\eta}]\subset W_{\zeta+\eta}$ and
 $[\hat{W}_{\zeta},\hat{W}_{\eta}]\subset \hat{W}_{\zeta+\eta}$.\label{binarysu8}}
\end{center}
\end{figure}

\newpage
\begin{figure}
\begin{center}
\[\begin{array}{c}
  \begin{array}{c}
  \hspace{-2pt}\sigma_{3}\otimes\hspace{2.2pt} I \\
  \hspace{3pt}I\hspace{2pt}\otimes\sigma_{3}  \\
  \sigma_{3}\otimes \sigma_{3}\\
  \end{array}
  \\
  \\
  \begin{array}{ccccc}
  W_{1}
  &
  \hspace{3pt}I\hspace{2pt}\otimes \sigma_{1}\hspace{15pt}\sigma_{3}\otimes \sigma_{1}
  &
  &
  \hspace{2pt}I\hspace{2pt}\otimes \sigma_{2}\hspace{13pt}\sigma_{3}\otimes \sigma_{2}
  &
  \hat{W}_{1}
  \\
  \\
  W_{2}
  &
  \hspace{-1pt}\sigma_{1}\otimes\hspace{3pt}I\hspace{15pt}\sigma_{1}\otimes \sigma_{3}
  &
  &
  \hspace{-1pt}\sigma_{2}\otimes\hspace{3pt} I\hspace{15pt}\sigma_{2}\otimes \sigma_{3}
  &
  \hat{W}_{2}
  \\
  \\
  W_{3}
  &
  \hspace{-1pt}\sigma_{1}\otimes \sigma_{1}\hspace{14pt}\sigma_{2}\otimes\sigma_{2}
  &
  &
  \hspace{-2pt}\sigma_{2}\otimes \sigma_{1}\hspace{13pt}\sigma_{1}\otimes \sigma_{2}
  &
  \hat{W}_{3}
 \end{array}
 \end{array}\]\\
  \fcaption{The quotient algebra of rank zero given by the intrinsic Cartan subalgebra of $su(4)$.\label{csu4}}
\end{center}
\end{figure}

\begin{figure}
\begin{center}
\[\begin{array}{c}
  \begin{array}{c}
  diag\{ 1,1,-1,-1 \}\\
  diag\{1,-1,1,-1 \}\\
  diag\{1,-1,-1,1 \}
  \end{array}
  \\
  \\
\begin{array}{ccccc}
 W_{01}
 &
 \lambda_{12}+\lambda_{34}\hspace{20pt}\lambda_{12}-\lambda_{34}
 &
 \hspace{40pt}
 &
 \hat{\lambda}_{12}+\hat{\lambda}_{34}\hspace{20pt}\hat{\lambda}_{12}-\hat{\lambda}_{34}
 &
 \hat{W}_{01}
 \\
 \\
 W_{10}
 &
 \lambda_{13}+\lambda_{24}\hspace{20pt} \lambda_{13}-\lambda_{24}
 &
 &
 \hat{\lambda}_{13}+\hat{\lambda}_{24}\hspace{20pt} \hat{\lambda}_{13}-\hat{\lambda}_{24}
 &
 \hat{W}_{10}
 \\
 \\
 W_{11}
 &
 \lambda_{14}+\lambda_{23}\hspace{20pt} \lambda_{14}-\lambda_{23}
 &
 &
 \hat{\lambda}_{14}+\hat{\lambda}_{23}\hspace{20pt}\hat{\lambda}_{14}-\hat{\lambda}_{23}
 &
 \hat{W}_{11}
 \end{array}
 \end{array}\] \fcaption{The quotient algebra of Fig.\protect\ref{csu4} in the $\lambda$-representation.\label{Gsu4}}
\end{center}
\end{figure}

\begin{figure}
\begin{center}
\[\begin{array}{c}
  \begin{array}{c}
   diag\{ 1,1,-1,-1 \} \\
   diag\{1,-1,1,-1 \} \\
   diag\{1,-1,-1,1 \}
  \end{array}
  \\
  \\
\begin{array}{ccccc}
 W_{01}
 &
 \lambda_{12}+\hat{\lambda}_{34}\hspace{20pt}\lambda_{12}-\hat{\lambda}_{34}
 &
 \hspace{40pt}
 &
 \hat{\lambda}_{12}+\lambda_{34}\hspace{20pt} \hat{\lambda}_{12}-\lambda_{34}
 &
 \hat{W}_{01}
 \\
 \\
 W_{10}
 &
 \lambda_{13}+\hat{\lambda}_{24}\hspace{20pt}\lambda_{13}-\hat{\lambda}_{24}
 &
 &
 \hat{\lambda}_{13}+\lambda_{24}\hspace{20pt}\hat{\lambda}_{13}-\lambda_{24}
 &
 \hat{W}_{10}
 \\
 \\
 W_{11}
 &
 \lambda_{14}+\hat{\lambda}_{23}\hspace{20pt}\lambda_{14}-\hat{\lambda}_{23}
 &
 &
 \hat{\lambda}_{14}+\lambda_{23}\hspace{20pt}\hat{\lambda}_{14}-\lambda_{23}
 &
 \hat{W}_{11}
 \end{array}
 \end{array}\] \fcaption{An alternative of the conjugate partition and the quotient algebra for that in Fig.\protect\ref{Gsu4}.\label{AGsu4}}
\end{center}
\end{figure}

\vspace*{-10pt} \noindent
\begin{thebibliography}{000}
\bibitem{OriginQAPSu} Z.-Y.~Su, {\it A Scheme of Cartan Decomposition for $su(N)$}, quant-ph/0603190.



\bibitem{Helgason}S. Helgason, {\em Differential geometry, Lie groups, and symmetric spaces},
 Academic Press, New York (1978).

\bibitem{Knapp}A.W. Knapp, {\em Lie groups beyond an introduction},
 Birkhauser, Boston (1996).

\bibitem{SuTsai1} Z.-Y. Su, {\it Quotient Algebra Partition and Cartan Decomposition for $su(N)$ II}.

\bibitem{SuTsai2} Z.-Y. Su and M.-C. Tsai, {\it Quotient Algebra Partition and Cartan Decomposition for $su(N)$ III}.

\bibitem{SuTsai3} Z.-Y. Su and M.-C. Tsai, {\it Quotient Algebra Partition and Cartan Decomposition for $su(N)$ IV}.

\bibitem{SuTsaiQAPFT} Z.-Y.~Su and M.-C.~Tsai,
{\it Every Action in Every Code is Fault Tolerant: Fault Tolerance
Quantum Computation in Quotient Algebra Partition}.

\bibitem{SuTsaiOptFTQC} M.-C.~Tsai and Z.-Y.~Su, {\it A Primitive Step Towards Optimized Fault Tolerance Quantum Computation}.


\bibitem{Khaneja1}N. Khaneja, R. Brockett, and S.J. Glaser, {\em Time optimal control
 in spin systems}, Phys. Rev. {\bf 63}, 032308 (2001).



\bibitem{Zhang} J. Zhang, J. Vala, K.B. Whaley, and S.
 Sastry, {\em  A geometric theory of non-local two-qubit
 operations}, Phys. Rev. A {\bf 67}, 042313 (2003). 

 \bibitem{SBullock}
 S. S. Bullock, G. K. Brennen and D. P. O'Leary,
 {\it Time Reversal and n-qubit Canonical Decompositions},
 Journal of Mathematical Physics, volume 46, 062104, 2005.



\end{thebibliography}
\end{document}